\documentclass[aps,prx,reprint,groupedaddress]{revtex4-2}

\usepackage{amsmath}
\usepackage{amsfonts}

\usepackage[colorlinks=true, pdfstartview=FitV, linkcolor=blue,
citecolor=blue, urlcolor=blue]{hyperref}
\usepackage[capitalise]{cleveref}

\usepackage{siunitx}

\usepackage{physics} 

\usepackage{graphicx}

\usepackage{xcolor}

\newcommand{\vect}[1]{\boldsymbol{#1}}

\begin{document}


\title{Enhanced orbital magnetic field effects in Ge hole nanowires}


\author{Christoph Adelsberger}
\author{Stefano Bosco}
\author{Jelena Klinovaja}
\author{Daniel Loss}
\affiliation{Department of Physics, University of Basel, Klingelbergstrasse 82, CH-4056 Basel, Switzerland}


\date{\today}

\begin{abstract}
Hole semiconductor nanowires (NW) are promising platforms to host spin qubits and Majorana bound states for topological qubits because of their strong spin-orbit interactions (SOI).
The properties of these systems depend strongly on the design of the cross section and on strain, as well as on external electric and magnetic fields.
In this paper, we analyze in detail the dependence of the SOI and $g$ factors on the orbital magnetic field. We focus on magnetic fields aligned along the axis of the NW, where orbital effects are enhanced and result in a renormalization of the effective $g$ factor up to $\SI{400}{\percent}$, even at small values of magnetic field.  
We provide an exact analytical solution for holes in Ge NWs and we derive an effective low-energy model that enables us to investigate the effect of electric fields applied perpendicular to the NW. We also discuss in detail the role of strain, growth direction, and high energy valence bands in different architectures, including Ge/Si core/shell NWs, gate-defined one-dimensional channels in planar Ge, and curved Ge quantum wells. By comparing NWs with different growth directions, we find that the isotropic approximation is well justified.
Curved Ge quantum wells feature large effective $g$ factors and 
	SOI at low electric field, ideal for hosting Majorana bound states. In contrast, at strong electric field, these quantities are independent of the field, making hole spin qubits encoded in curved quantum wells to good approximation not susceptible to charge noise, and significantly boosting their coherence time.

\end{abstract}


\maketitle

\section{Introduction}

Semiconducting nanostructures based on holes are emerging as frontrunner candidates to process quantum information because of their large spin-orbit interaction (SOI)~\cite{Froning2021,Hao2010,Hu2011,Terrazos2021,Scappucci2021,Liu2022} that enables ultrafast and coherent manipulations of spin qubits~\cite{Hendrickx2020,Hendrickx2021,Watzinger2018,Jirovec2021,Geyer2021,Qvist2022a}, strong coupling to resonators~\cite{Kloeffel2013,Bosco2022a,Michal2022}, and is an essential ingredient to host exotic particles such as Majorana bound states (MBSs)~\cite{Kitaev2001,Klinovaja2012}.
In hole nanostructures, the SOI is not only surprisingly strong, orders of magnitude larger than in electronic systems~\cite{Froning2021,Froning2021a,Wang2022},  but it is also highly tunable by external electromagnetic fields and it can be engineered by the confinement potential and by strain~\cite{Bulaev2005,Bulaev2007,Kloeffel2011,Kloeffel2014,Kloeffel2018,Bosco2021,Philippopoulos2020,Hu2007,Adelsberger2022}, resulting in sweet spots where the charge noise plaguing state-of-the-art spin qubits is strongly suppressed~\cite{Bosco2021b,Wang2021,Bosco2022}. The qubit coherence is further enhanced by the weak hyperfine noise, another crucial issue for spin-based quantum information processing~\cite{Witzel2006,Yao2006,Cywinski2009,Khaetskii2002,Coish2004,Hanson2007}, that in hole spin qubits encoded in Si and Ge quantum dots (QDs) can be suppressed by isotopic purification~\cite{Itoh1993,Itoh2014} or by an appropriate  QD design~\cite{Fischer2010,Maier2012,Klauser2006,Prechtel2016,Testelin2009,Fischer2008, Bosco2021a}.

In particular, the largest SOI arises in quasi-one-dimensional architectures, including Ge nanowires (NWs)~\cite{Kloeffel2018, Watzinger2018,Froning2021,Milivojevi2021} and gate-defined squeezed QDs in planar heterostructures~\cite{Bosco2021}.
In these systems, experiments have shown a large proximity-induced superconductivity~\cite{Doh2005,Xiang2006,Hajer2019}, making hole NWs promising candidates for topological quantum information processing based on MBSs~\cite{Alicea2012,Maier2014,Albrecht2016,Lutchyn2010,Lutchyn2018,Mourik2012,Sau2010,Oreg2010,Alicea2011,Mao2012,Gangadharaiah2011}. 
A stable topological phase, however, also requires a large $g$ factor, that allows to reach a sufficiently large Zeeman energy overcoming the induced superconducting gap even at the weak magnetic fields compatible with superconductors~\cite{Alicea2011,Laubscher2021,Klinovaja2012,Maier2014,Dmytruk2018}.

Orbital magnetic field effects play a crucial role in defining the property of hole nanostructures, yielding significant corrections of the $g$ factor and of the effective mass in planar heterostructures~\cite{Stano2018, Stano2019} as well as  in NWs~\cite{Adelsberger2022,Bosco2022,Legg2021}. Orbital effects are also used to study the shape anisotropy in gate defined quantum dots~\cite{Camenzind2021}. 
In hole NWs, these effects are enhanced by magnetic fields that point along the direction of the NW,  where we will show that they yield a renormalization of the $g$ factor as large as $400\%$. 

\begin{figure}[htb]
	\includegraphics{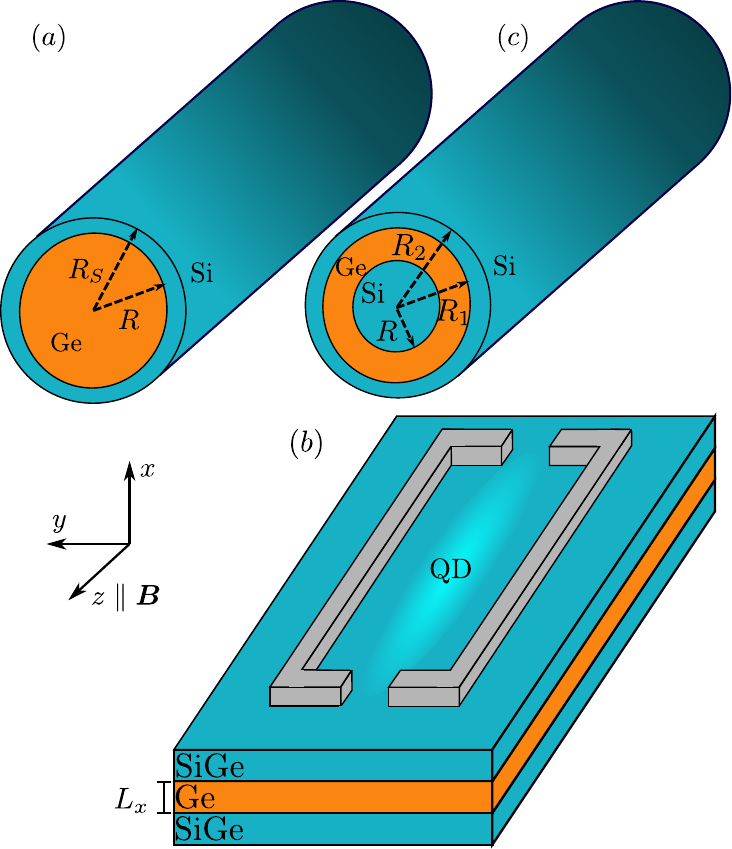}
	\caption{Sketches of the architectures analyzed. (a) Cylindrical NW with radius $R$ extending along the $z$ axis. The NW core (Ge) is covered by a shell (Si) that induces strain into the core. We denote by $R_S$ the radius of the NW including the shell. (b) Sketch of a planar Ge/SiGe heterostructure with gate-defined one-dimensional channel. The Ge layer has width $L_x$ and the channel is electrostatically defined by a harmonic confinement in the $y$ direction with harmonic length $l_y$. (c) Sketch of a CQW: a NW with an annular cross section. The setup consists of a Si core of radius $R$ covered by an inner Ge shell (radius $R_1$) that hosts the holes. Another Si layer (radius $R_2$) covers the Ge shell. Strain is induced into the Ge layer from the Si core and the outer Si shell. In all three cases we consider a magnetic field of strength $B = \abs{\vect{B}}$ applied parallel to the $z$ axis as indicated by the coordinate system. \label{fig:Three1dSystems}} 
\end{figure}

In this paper, we demonstrate the importance of orbital magnetic field effects in one-dimensional hole systems in Ge, see Fig.~\ref{fig:Three1dSystems}. We present an analytical solution, exactly including orbital effects, for low-energy holes in isotropic semiconductor NWs in the presence of a magnetic field parallel to the NW axis. This solution allows us to derive an effective low-energy model describing the effect of homogeneous and inhomogeneous electric fields perpendicular to the NW. Our model unravels the strong dependence of the $g$ factor, SOI, and effective mass on external electromagnetic fields and on strain. We discuss the emergence of a spin-dependent mass term, that appears also at magnetic fields perpendicular to the NW axis~\cite{Adelsberger2022}. 

Moreover, in Si/Ge core/shell NWs,  strain is crucial to increase the subband gap between the lowest pair of energy states and the excited states~\cite{Kloeffel2018,Kloeffel2014,Adelsberger2022}. 
We analyze its effect analytically and predict that strain enhances the $g$ factor at the cost of weaker SOI. 
This effect is understood by introducing a strain-induced energy scale which in these systems favors a light hole (LH) ground state over mixed heavy hole (HH)-LH states.

We compare our analytical results for Ge NWs to numerical calculations of gate-defined one-dimensional channels in a planar heterostructure~\cite{Adelsberger2022,Bosco2021} and curved Ge quantum wells (CQW)~\cite{Bosco2022}.
Gate-defined channels exhibit a smaller $g$ factor and weaker SOI than in core/shell NWs, but with a similar qualitative behavior of the effective parameters against electric field and strain. 
In contrast, the response of the CQWs is strikingly different. 
In these architectures, the orbital magnetic field effects are extremely important because of their annular cross section, yielding  large $g$ factors and  strong SOI at weak electric field, ideal to host MBSs. 
Moreover, we show that the $g$ factor and the SOI remain constant in a wide range of strong electric fields, and thus CQWs are an ideal platform to encode spin qubits, with a strongly suppressed susceptibility to charge noise, a crucial issue in core/shell NWs~\cite{Froning2021a,Wang2022}.

Furthermore, we numerically analyze corrections to our model, focusing in particular on the influence of holes belonging to the spin-orbit split-off band (SOB) and on cubic anisotropies of the valence band, which are neglected in our analytical calculations. The SOB does not qualitatively alter our predictions and only changes the quantitative values of the parameters. Most notably, in narrow  NWs, the SOB reduces the $g$ factor significantly; the SOI is less affected by the SOB. Our analysis of anisotropies reveals that the isotropic approximation (IA) is well justified for the description of NWs grown along one of the main crystallographic axes, along the  $[110]$ (often used in experiments~\cite{Voisin2015, Maurand2016, Barraud2012}), or along the $[111]$ direction. This result further justifies the application of the IA in the rest of the paper.

This paper is organized as follows. In Sec.~\ref{sec:ModelOfNW} we introduce the model used to describe low-energy holes in semiconductor NWs. Our analytical solution for a NW with magnetic field parallel to the NW taking orbital effects of the magnetic field exactly into account is derived and discussed in Sec.~\ref{sec:AnaSolNW}. We start with briefly showing the non-parabolic bulk dispersion relation in the presence of orbital effects and continue by adding hard-wall (HW) boundary conditions to model NWs with circular cross sections. Furthermore, we investigate  the effective $g$ factor, the direct Rashba SOI~\cite{Kloeffel2011,Kloeffel2018}, and the effective mass of the holes by considering an effective low-energy model to second order perturbation theory, also including homogeneous and inhomogeneous electric fields perpendicular to the NW. Moreover, we discuss how strain increases the $g$ factor, the subband gap between ground state and excited states, and the effective mass, while it decreases the SOI and magnitude of the spin-dependent mass. In Sec.~\ref{sec:1dChannel} we compare our results to numerically calculated effective parameters of a gate-defined one-dimensional channel defined in planar heterostructures. Likewise we study numerically holes in CQWs consisting of a Ge shell that hosts the holes around a Si core in Sec.~\ref{sec:thinShell}. In Sec.~\ref{sec:corrections} we extend our model for the NW by correcting terms and analyze its validity. In particular, we include the SOB and find that these states cause quantitative corrections to the effective parameters. Finally, we investigate the anisotropies in Ge/Si core/shell and CQWs. Conclusion and outlook are provided in Sec.~\ref{sec:conclusion}. Details on the calculations are given in the Appendices.

\section{Model of Nanowire \label{sec:ModelOfNW}}
In this section we present the theoretical model used in this paper.
The general Hamiltonian modeling hole nanostructures  in the presence of a magnetic field is given by
\begin{align}
H = H_\mathrm{LK}+H_Z + H_\mathrm{BP}   + V \ . \label{eqn:FullHamiltonian}
\end{align}

The Luttinger-Kohn (LK) Hamiltonian  $ H_\mathrm{LK}$ describes the mixing of HHs and LHs, and by including orbital magnetic field effects it is given by~\cite{Luttinger1955,Luttinger1956,Kloeffel2018,Lipari1970,Winkler2003}
\begin{align}
H_\text{LK} = &\frac{\hbar^2}{2 m} \left[\gamma_k \pi^2 -2 \gamma_2 (\pi_{x'}^2J_{x'}^2 + \pi_{y'}^2 J_{y'}^2 +\pi_{z'}^2 J_{z'}^2)\right.\nonumber\\
&\left. \vphantom{\pi^2_x}\! - 4 \gamma_3 \left(\{\pi_{x'}, \pi_{y'}\}\{J_{x'}, J_{y'}\} + \text{c.p.}\right)\right], \label{eqn:LK_Hamiltonian}
\end{align}
where $\{A, B\}=(A B+ B A)/2$, $\gamma_k = \gamma_1 + 5\gamma_2/2$, and  ``c.p.'' means cyclic permutations. The primed indices $ x'$, $y'$, and  $z'$ denote the axes aligned to the main crystallographic axes $[100]$, $[010]$, and $[001]$, respectively, and $J_i$ [with $i = x', y', z'$] are the standard spin-$3/2$ operators. The HHs correspond to the spin component $\pm3/2$ and the LHs to $\pm1/2$ of $J_z$. The coefficients $\gamma_1$, $\gamma_2$, and $\gamma_3$ are the material-dependent Luttinger parameters~\cite{Winkler2003} and $m$ is the bare electron mass. Furthermore, the kinematic momentum operators in Eq.~\eqref{eqn:LK_Hamiltonian} include orbital effects of the magnetic field by the Peierls substitution~\cite{Luttinger1956}
\begin{align}
\vect{\pi} = \vect{k} + \frac{e}{\hbar} \vect{A} \label{eqn:kinElMom}
\end{align} 
with canonical momenta $k_j = - i \hbar \partial_j$,  positive elementary charge $e>0$, and vector potential $\vect{A}$. 
The magnetic field $\vect{B}$ also splits the spin states by the  Zeeman Hamiltonian
\begin{align}
	H_Z = 2 \kappa \mu_B \vect{B}\cdot \vect{J} \ ,\label{eqn:HamZeeman}
\end{align}
with $\kappa=3.41$ in Ge. We neglect here the small anisotropic Zeeman energy $\propto J_i^3$~\cite{Luttinger1956, Kloeffel2018}.
In this paper we focus on  magnetic fields aligned to the NW; a detailed analysis of perpendicular magnetic fields is provided in Ref.~\cite{Adelsberger2022}. 

In Ge the Luttinger parameters $\gamma_1 = 13.35$, $\gamma_2 = 4.25$, and $\gamma_3 = 5.69$~\cite{Lawaetz1971} describe a relatively isotropic semiconductor with $(\gamma_3-\gamma_2)/\gamma_1\approx 0.1$, thus,  we will often use the approximate isotropic Luttinger-Kohn (ILK) Hamiltonian
\begin{align}
	H_\mathrm{ILK} = \frac{\hbar^2}{2 m} \left[\gamma_k \pi^2 - 2 \gamma_s (\vect{k} \cdot \vect{J})^2\right] + H_\mathrm{orb}, \label{eqn:LK_Hamiltonian_sphericalApp}
\end{align}
commonly adopted in literature~\cite{Kloeffel2018,Kloeffel2011,Bosco2021,Lipari1970,Li2021,Li2021b}, with 
\begin{align}
	&H_\mathrm{orb}  =\frac{\hbar e}{2 m} \bigg\{\gamma_k \left(\frac{e}{\hbar}\vect{A}^2 + 2\vect{k}\cdot \vect{A}\right) - \frac{2 \gamma_s  e}{\hbar}\left(\vect{A}\cdot\vect{J}\right)^2\nonumber\\
	&-4\gamma_s \left[k_x A_x J_x^2  + \left(\left\{k_x, A_y\right\} + \left\{k_y, A_x\right\}\right) \left\{J_x, J_y\right\}+ \mathrm{c.p.}\right]\bigg\},
\end{align}
where $\gamma_s = (\gamma_2 + \gamma_3)/2 = 4.97$.
We stress that while our analysis here is restricted to Ge, our analytical results are valid more generally for  holes in GaAs, InAs, or InSb, where the ILK is valid~\cite{Winkler2003}.

The quasi-one-dimensional system is defined by a confinement potential $V$ that models the different NWs schematically depicted in Fig.~\ref{fig:Three1dSystems}. 
In this paper, we first consider a Ge/Si core/shell NW with cylindrical cross section [Fig.~\ref{fig:Three1dSystems}(a)]. 
The Ge NW of radius $R$ is covered by a Si shell of thickness $R_S-R$, that produces a large strain on the Ge core. This strain is known to play a relevant role for the properties of the NW~\cite{Kloeffel2014} and we include its effect by the Bir-Pikus (BP) Hamiltonian~\cite{Bir1974}. In this setup, the BP Hamiltonian is well-approximated by~\cite{Kloeffel2018} 
\begin{align}
	H_\textrm{BP} &= \abs{b} \varepsilon_s  J_{z'}^2. \label{eqn:BPHam}
\end{align} 
The strain energy $\abs{b} \varepsilon_s$, with $\varepsilon_s = \varepsilon_{\perp}  - \varepsilon_{z'z'}>0$, is typically positive~\cite{Kloeffel2014} and it comprises the deformation potential $b=\SI{-2.5}{\electronvolt}$~\cite{Bir1974} and the strain tensor elements $\varepsilon_{\perp}$ and $\varepsilon_{z'z'}$. Under the assumption of homogeneous strain in the core of the NW, these strain elements depend only on the relative shell thickness $\gamma = (R_S-R)/R$~\cite{Kloeffel2014, Menendez2010}.  For the typical value $\gamma = 0.1$, one finds~\cite{Kloeffel2014} $\abs{b} \varepsilon_s = \SI{15.5}{\milli\electronvolt}$.

For the gate-defined one-dimensional channel in a planar Ge/SiGe heterostructure as depicted in Fig.~\ref{fig:Three1dSystems}(b), we consider a Ge layer with width $L_x$ and confined in the $y$ direction by a harmonic potential parametrized by the harmonic length $l_y$. This setup describes squeezed quantum dots in planar Ge~\cite{Adelsberger2022,Bosco2021}. 
In this case, the strain due to the lattice mismatch between the Ge and SiGe layers results in the BP Hamiltonian~\cite{Terrazos2021,Wang2021}
\begin{align}
	H_\mathrm{BP}^\mathrm{ch} = \abs{b} \varepsilon_s J_x^2. \label{eqn:BPHam_channel}
\end{align}
In contrast to the core/shell NW case [cf. Eq.~\eqref{eqn:BPHam}] the strain energy $\abs{b}\varepsilon_s<0$, and  the strain favors a HH groundstate, with quantization axis perpendicular to the substrate. The strain energy can be engineered by the percentage of Si in the SiGe layers.

Moreover, we compare Ge/Si core/shell NWs to a CQW sketched in Fig.~\ref{fig:Three1dSystems}(c). The CQW consists of a Si core of radius $R$, a thin Ge shell of thickness $R_1-R$ that hosts the holes, and an outer Si shell of thickness $R_2-R_1$.  In addition to the longitudinal strain typical of core/shell NWs [see Eq.~\eqref{eqn:BPHam}], the CQW is also subject to a radial strain that resembles the strain in planar heterostructures. Explicitly, the total BP Hamiltonian  of CQWs is well-approximated by~\cite{Bosco2022}
\begin{align}
	H_\mathrm{BP}^\mathrm{CQW} = \abs{b}\left( \varepsilon_z J_z^2 - \varepsilon_r J_r^2 \right) \label{eqn:HBP_ts}
\end{align}
where we define the radial spin-$3/2$ matrix as
\begin{align}
	J_r = \vect{\hat{e}_r} \cdot \vect{J} , 
\end{align}
with the unit vector in radial direction $\vect{\hat{e}_r} = (\cos\theta, \sin\theta,0 )$. In Eq.~\eqref{eqn:HBP_ts} we approximate the longitudinal and radial strain energies as~\cite{Bosco2022}
\begin{align}
	\varepsilon_z &\approx \varepsilon_p \frac{R_1-R}{R_1+R} \left(1 - \frac{R_1-R}{2(R_1+R)}-\frac{(R_1+R)^2}{2 R_2^2}\right) , \label{eqn:strainLong} \\
	\varepsilon_r &\approx \varepsilon_p \left(1 - \frac{R_1-R}{R_1+R}\right)^2,\label{eqn:strainRad}
\end{align}
respectively.
We assume here $\abs{b} \varepsilon_p \approx \SI{140.8}{\milli\electronvolt}$, a value which can be reduced by replacing the Si in the inner and outer shells  by a  Si$_x$Ge$_{1-x}$ alloy.

In order to study the validity of the ILK approximation in Eq.~\eqref{eqn:LK_Hamiltonian_sphericalApp}, we examine also the effect of the cubic anisotropies of the LK Hamiltonian in Eq.~\eqref{eqn:LK_Hamiltonian}. In general, the behavior of the system depends on the growth direction of the NW. In this paper, we focus on a few important cases.
Particularly relevant examples are NWs grown along the $[001]$ axis, where the SOI is maximized or fully tunable ~\cite{Kloeffel2018,Bosco2021b}.
We also study NWs grown along the [110] crystallographic axis, consistent with several recent experiments~\cite{Voisin2015, Maurand2016, Barraud2012}.
Finally, we consider another relevant growth direction that is experimentally achievable~\cite{McIntyre2020,ConesaBoj2017,Pei2010,Xiang2010,Adhikari2006,Jagannathan2006}, where the NW is grown along the $z\parallel [111]$ direction.
The explicit form of the rotated LK Hamiltonian in these cases is reported in Ref.~\cite{Adelsberger2022}.

\section{Analytical Solution for Ge Nanowires} \label{sec:AnaSolNW}

In the following we utilize the ILK Hamiltonian from Eq.~\eqref{eqn:LK_Hamiltonian_sphericalApp} to find an exact analytical description of low-energy holes in a Ge or a Ge/Si core/shell NW. 
Our approach fully accounts for orbital magnetic field effects that are typically neglected or included perturbatively~\cite{Nowack2007,Li2021,Li2021b,Qvist2022}. We find that, in Ge, these effects strongly  renormalize the system response, and, thus, cannot be safely neglected even at weak magnetic fields. With the help of our analytical results, we analyze in detail the behavior of the system under the effect of both, electric and magnetic fields, and we include also strain in Ge/Si core/shell NWs.

To describe the analytical procedure, we begin with the analysis of the ILK Hamiltonian, cf. Eq.~\eqref{eqn:LK_Hamiltonian_sphericalApp}, in the bulk in Sec.~\ref{sec:BulkSolution}. Therefore, we separate the ILK Hamiltonian into three parts according to their order in the momentum $\pi_z$~\cite{Adelsberger2022}
\begin{align}
	H_\mathrm{ILK} = H_{xy} + H_\mathrm{int} \pi_z + H_{zz} \pi_z^2. \label{eqn:Factorization}
\end{align}
First, we present a simple approach where we derive the bulk solution for $H_{xy}$ where the magnetic field is taken into account exactly since it is the starting point for the perturbation theory in $\pi_z$. However, in the bulk, we do not resort to perturbation theory but give an exact solution for the dispersion relation in Appendix~\ref{sec:BulkDispersionRelation}. In Sec.~\ref{sec:NWAnalyticalSolution}, we proceed for a NW in analogy to the bulk by deriving an exact solution for $H_{xy}$ including the orbital effects of the magnetic field exactly. This can be highly relevant because the effective $g$ factor is considerably renormalized due to orbital effects even at weak magnetic field as discussed in Sec.~\ref{sec:orbitalCorrections}. In the NW it is not as straightforward as in the bulk to find an exact solution for finite $\pi_z$. Thus, we develop an effective theory in perturbation theory in Sec.~\ref{sec:EffHam} and Sec.~\ref{sec:2x2wireHam} following Ref.~\cite{Kloeffel2018}, where we discuss the effects of homogeneous and inhomogeneous electric fields. Finally, in Sec.~\ref{sec:strain} we discuss strain in Ge/Si core/shell NWs.

\subsection{Bulk solution \label{sec:BulkSolution}}
For the following derivation of the bulk solution we assume a magnetic field along the $z$ direction that we will identify with the NW axis in Sec.~\ref{sec:NWAnalyticalSolution}.
The complete analytical solution of the bulk ILK Hamiltonian in this case is provided in Appendix~\ref{sec:BulkDispersionRelation}, but the solution is rather complicated and not straightforwardly generalizable to NWs.

Instead, in this section we discuss a simpler approach, that will provide the starting point for our analysis of NWs.
We restrict ourselves to the analysis of long wave length excitations, with a wave length $1/k_z$ much longer than the characteristic length defining the variation of the wave function in the $x,y$ plane. In the bulk analysis, this condition means that $k_z l_B \ll 1$, where $l_B=\sqrt{\hbar/eB}$ is the magnetic length. 
At small values of $k_z$, the system is well described by the eigenstates of $H_{xy}$, and the $k_z$ dependence can be studied by treating $H_\mathrm{int}$ and $H_{zz}$ in perturbation theory. 

To better study the magnetic orbital effects, we now neglect the Zeeman energy. To do so, we introduce the Landau ladder operator
\begin{align}
	a =\frac{\pi_x - i \pi_y}{\sqrt{2}}l_B, \label{eqn:LandauLevelOp}
\end{align}
 obeying the canonical commutation relation $\left[a, a^\dagger\right] = 1$. In the spin basis ($+3/2$, $-1/2$, $-3/2$, $+1/2$), $H_{xy}$ becomes 
\begin{align}
	&\frac{H_{xy}}{\hbar \omega_c} =\begin{pmatrix}
		\gamma_+ \left( a^\dagger a + \frac{1}{2}\right) & - \sqrt{3} \gamma_s a^2 \\ 
		- \sqrt{3} \gamma_s \left(a^\dagger\right)^2 & \gamma_- \left( a^\dagger a + \frac{1}{2}\right)
	\end{pmatrix}   \nonumber\\
&\hspace{50pt}	\oplus \begin{pmatrix}
		\gamma_+ \left( a^\dagger a + \frac{1}{2}\right) & - \sqrt{3} \gamma_s \left(a^\dagger\right)^2 \\ 
		- \sqrt{3} \gamma_s a^2 & \gamma_- \left( a^\dagger a + \frac{1}{2}\right)
	\end{pmatrix}, \label{eqn:HamPerp}
\end{align}
where $\gamma_\pm = \gamma_1 \pm \gamma_s$ and the symbol ``$\oplus$'' refers to the direct sum of matrices. The energy is given here in units of $\hbar \omega_c$, with cyclotron frequency $ \omega_c= e B/m$, and the lengths are in units of $l_B$. Focusing on the upper block ($\uparrow$) and solving the Schr\"{o}dinger equation
\begin{align}
\frac{1}{\sqrt{3} \gamma_s} \left[\gamma_+ \left(a^\dagger a + \frac{1}{2}\right) - \varepsilon\right] \psi_\mathrm{HH}&= a^2 \psi_\mathrm{LH},\\
\frac{1}{\sqrt{3} \gamma_s} \left[\gamma_- \left(a^\dagger a + \frac{1}{2}\right) - \varepsilon\right] \psi_\mathrm{LH}&= \left(a^\dagger\right)^2 \psi_\mathrm{HH},
\end{align} 
with the HH (LH) components of the wave function $\psi_\mathrm{HH (LH)}$ we find for the energy spectrum written in magnetic units
\begin{align}
	&\varepsilon_\pm^\uparrow(\bar{n}) = \gamma _1 \left(\bar{n}-\frac{1}{2}\right)\nonumber\\
	&\hspace{5pt}\pm\sqrt{\gamma _1^2+  (1-2\bar{n})\gamma _1 \gamma _s+\left[4\bar{n} (\bar{n}-1)+\frac{1}{4}\right] \gamma _s^2}, \label{eqn:bulkEnergy}
\end{align}
where $\bar{n}$ is the eigenvalue of the number operator $a^\dagger a$. In the bulk solution, $\bar{n}$ is an integer because the wave function is required to vanish  at infinity. In the absence of magnetic fields, time-reversal symmetry implies that for the energy spectrum of the lower block ($\downarrow$) is the same as for $\uparrow$.

Starting from these solutions for $\pi_z=0$, we can proceed with a perturbation theory in $\pi_z$ to find the dispersion relation for the different Landau level subbands. We will do such a perturbation theory in Secs.~\ref{sec:EffHam} and~\ref{sec:2x2wireHam} for the NW. However, in the bulk case it is possible to derive an exact analytical solution that we provide in Appendix~\ref{sec:BulkDispersionRelation}.

\begin{figure}[tb!]
	\includegraphics{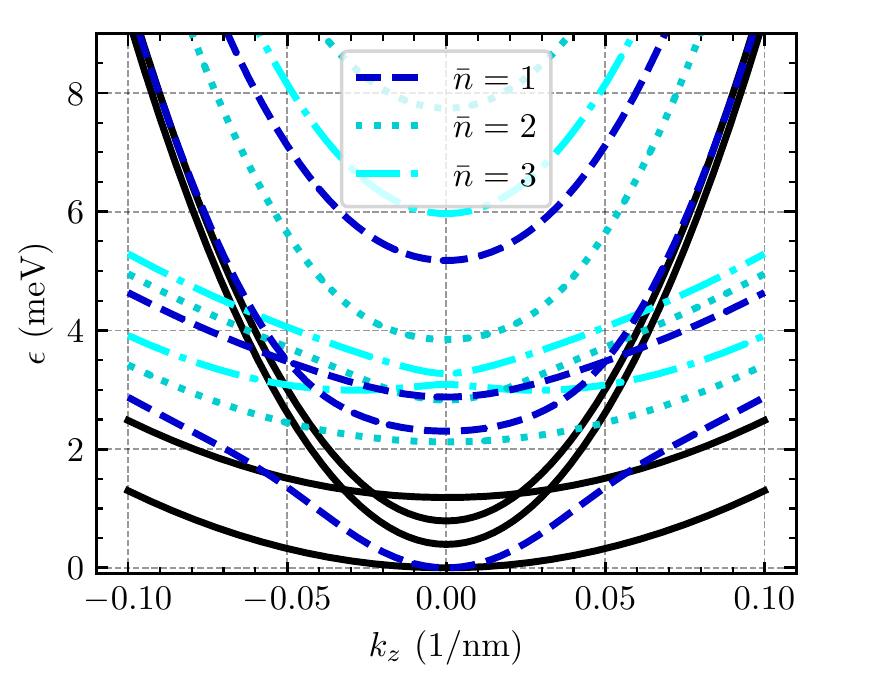}
	\caption{Comparison of the bulk dispersion relations of holes in Ge with (color lines) and without (black solid lines) orbital effects according to the analytical solution in Appendix~\ref{sec:BulkDispersionRelation}. Eq.~\eqref{eqn:bulkEnergy} gives the bulk energies of the upper block at $k_z=0$ and $B=0$. For the curves that include orbital effects, we consider $B=\SI{1}{\tesla}$ and we depict the dispersion relation for the lowest three $\bar{n}$ as indicated in the legend. Orbital effects modify the dispersion relation, yielding even a non-parabolic dispersion relation for the ground state. For the curves without orbital effects we choose $k_x =k_y = 0$. \label{fig:bulkSpec_B1}}
\end{figure}

In Fig.~\ref{fig:bulkSpec_B1} we compare the bulk dispersion relations of holes in Ge with and without orbital effects given in Appendix~\ref{sec:BulkDispersionRelation}. The figure illustrates how the dispersion relation deviates from a parabolic spectrum (black lines) in the presence of orbital effects (color lines). Importantly, we stress that the orbital effects strongly renormalize the mass and even result in a non-parabolic dispersion of the ground state, as we can observe by comparing to the ground state without orbital magnetic fields (black lines). 
\\

\subsection{Cylindrical nanowire with hard-wall confinement \label{sec:NWAnalyticalSolution}}
In this subsection, we consider a cylindrical NW with radius $R$ in a magnetic field $\vect{B}= (0,0,B)$ in the $z$ direction  parallel to the NW, see Fig.~\ref{fig:Three1dSystems}(a). For now, we neglect the strain present in Ge/Si core/shell NWs and we will include it in Sec.~\ref{sec:strain}.  A convenient gauge  in this case is the symmetric gauge $\vect{A}= (-y, x, 0) B/2$, which preserves the rotational invariance of the cross section. Thus the total angular momentum $I_z = s_z + L_z$, with effective spin $s_z$ and orbital angular momentum $L_z$, is preserved too.
Moreover, as $\vect{B}\parallel z$, the translational invariance along the NW is also preserved and $\pi_z=k_z$ is a good quantum number (QN). We assume a hard-wall (HW) confinement potential
\begin{align}
V(r) = 
\begin{cases}
0, &r = \sqrt{x^2 + y^2}  < R,\\ 
\infty, &\mathrm{otherwise.}
\end{cases} \label{eqn:HW_boundaryCondition}
\end{align}
Following the procedure described in Sec.~\ref{sec:BulkSolution}, we find that at $k_z=0$ the unnormalized wave function of the $\uparrow$ block of the Hamiltonian in Eq.~\eqref{eqn:HamPerp} is given in the presence of the HW boundary condition [see Eq.~\eqref{eqn:HW_boundaryCondition}] by
\begin{align}
&\Psi_{m}^\uparrow= \begin{pmatrix}
\Psi_{m}^{+3/2}\\
\Psi_{m}^{-1/2}
\end{pmatrix}  
= L_{\alpha_-^\uparrow}^m \left(\frac{R^2}{2}\right) 
\begin{pmatrix}
\psi_{m, \alpha_+^\uparrow} \left(\vect{r}\right)\\
c_+^\uparrow \psi_{m-2, \alpha_+^\uparrow+2} \left(\vect{r}\right)
\end{pmatrix}\nonumber\\
& \hspace{30pt}- L_{\alpha_+^\uparrow}^m \left(\frac{R^2}{2}\right)
\begin{pmatrix}
	\psi_{m, \alpha_-^\uparrow} \left(\vect{r}\right)\\
	c_-^\uparrow	\psi_{m-2, \alpha_-^\uparrow+2} \left(\vect{r}\right)
\end{pmatrix}, \label{eqn:WF_GeNWU}
\end{align}
where the first component of the spinor, $\Psi_{m}^{+3/2}$, corresponds to the HH spin $+3/2$ and the second,  $\Psi_{m}^{-1/2}$, to the LH spin $-1/2$. Here, $L_{\alpha}^m \left(x\right)$ is the associated Laguerre function and 
\begin{align}
	\psi_{m, \alpha} (\vect{r}) = {i^m}2^{-\frac{m}{2}} e^{-im\varphi} e^{-\frac{r^2}{4}}r^m L^m_{\alpha}\left(\frac{r^2}{2}\right),
\end{align}
where $\vect{r}$ is given in polar coordinates with the radial coordinate $r$ and the angle $\varphi$. A detailed derivation of this result is provided in Appendix~\ref{sec:AnalyticsDerivationNW}. Note that all lengths are given in units of $l_B$.  In analogy to the solution in Sec.~\ref{sec:BulkSolution}, the  spin QN is now substituted by a pseudo-spin that we denote by $\uparrow$ and $\downarrow$ referring to the two blocks of $H_{xy}$ in Eq.~\eqref{eqn:HamPerp}. The additional QN $m$ is an integer that is related to the total angular momentum QN $I_z$ by $m = 2 I_z+1$. The coefficients $	\alpha_{\pm}^\uparrow $ are given by
\begin{widetext}
\begin{align}
	\alpha_{\pm}^\uparrow = &\Bigg\{-\frac{3}{2} \gamma_1^2 +\gamma_1 (\varepsilon -\frac{\kappa}{2} )+ \gamma_s (6 \gamma_s+\kappa ) \nonumber\\
	&\pm\bigg[\gamma_1^2 \left( \gamma_1^2-\frac{23}{4}\gamma_s^2-2\gamma_s\varepsilon\right)- \kappa  (\gamma_1-2 \gamma_s) \left[2 \gamma_1^2+3 \gamma_1 \gamma_s-2 \gamma_s (\varepsilon +\gamma_s)\right] \nonumber\\
	&+ \kappa ^2 (\gamma_1+\gamma_s) (\gamma_1-2 \gamma_s)+  \gamma_s^2(7\gamma_s^2+8\gamma_s\varepsilon + 4\varepsilon^2)\bigg]^{1/2}\Bigg\}/(\gamma_1^2-4 \gamma_s^2), \label{eqn:alphaU}
\end{align}
\end{widetext}
which are real numbers and depend on the Zeeman energy via $\kappa$. If $\alpha_\pm^s$, $s = \uparrow, \downarrow$, is an integer and $\kappa=0$, we recover the bulk solution in Eq.~\eqref{eqn:bulkEnergy}. The coefficients $ c_{\pm}^\uparrow$ are given by
\begin{align}
 	c_{\pm}^\uparrow= \frac{(2 \alpha_{\pm}^\uparrow +1) (\gamma_1+ \gamma_s)+3\kappa -2 \varepsilon }{2 \sqrt{3} \gamma_s}.  \label{eqn:cU}
\end{align}
Imposing the HW boundary conditions, which follow from  Eq.~\eqref{eqn:HW_boundaryCondition}, on the wave functions defined in Eq.~\eqref{eqn:WF_GeNWU}, we find  the implicit eigenvalue equation determining the energy $\varepsilon_m^s(n)$,
\begin{align}
\frac{c_-^\uparrow}{c_+^\uparrow}=\frac{L^m_{\alpha_-^\uparrow}\left(\frac{R^2}{2}\right)
	L^{m-2}_{\alpha_+^\uparrow +2}\left(\frac{R^2}{2}\right)}{L^m_{\alpha_+^\uparrow}\left(\frac{R^2}{2}\right)
	L^{m-2}_{\alpha_-^\uparrow +2}\left(\frac{R^2}{2}\right)}. \label{eqn:dispRelGeU}
\end{align}
Here we introduce an additional QN $n$ to number the energies consecutively by their magnitude for each $m$. 
In analogy, for the $\downarrow$ block of $H_{xy}$ in Eq.~\eqref{eqn:HamPerp} describing the spin states ($-3/2$, $+1/2$) we find the spinor
\begin{align}
	&\Psi_{m}^\downarrow = \begin{pmatrix}
		\Psi_{m}^{-3/2}\\
		\Psi_{m}^{+1/2}
	\end{pmatrix}  
	= L_{\alpha_-^\downarrow}^{-m} \left(\frac{R^2}{2}\right) 
	\begin{pmatrix}
		c_+^\downarrow \psi_{m-2, \alpha_+^\downarrow+2} \left(\vect{r}\right)\\
		\psi_{m, \alpha_+^\downarrow} \left(\vect{r}\right)
	\end{pmatrix}\nonumber\\
	&\hspace{20pt}- L_{\alpha_+^\downarrow}^{m} \left(\frac{R^2}{2}\right)
	\begin{pmatrix}
		c_-^\downarrow\psi_{m-2, \alpha_-^\downarrow+2}  \left(\vect{r}\right)\\
		\psi_{m, \alpha_-^\downarrow}  \left(\vect{r}\right)
	\end{pmatrix}.\label{eqn:WF_GeNWD}
\end{align}
Together with the solution for the $\uparrow$ block this is the exact analytical solution for an isotropic semiconductor hole NW with circular cross section in a magnetic field parallel to the NW axis. The coefficients $\alpha_\pm^\downarrow$ and $	c_\pm^\downarrow$ for the $\downarrow$ block are
\begin{widetext}
\begin{align}
		\alpha_\pm^\downarrow =&\Bigg\{-\frac{3}{2} \gamma_1^2 +\gamma_1 (\varepsilon +\frac{\kappa}{2})+\gamma_s(6 \gamma_s - \kappa)\nonumber\\ 
		&\pm\bigg[\gamma_1^2 \left(\gamma_1^2 - \frac{23}{4} \gamma_s^2+2 \gamma_s\varepsilon\right)- \kappa  (\gamma_1-2 \gamma_s) \left[2 \gamma_1^2+3 \gamma_1 \gamma_s+2 \gamma_s (\varepsilon -\gamma_s)\right]\nonumber\\
			&+\kappa ^2 (\gamma_1+\gamma_s) (\gamma_1-2 \gamma_s)+ \gamma_s^2 (7 \gamma_s^2 - 8\gamma_s \varepsilon +4 \varepsilon^2)\bigg]^{1/2}\Bigg\}/(\gamma_1^2-4 \gamma_s^2) \label{eqn:alphaD}
\end{align}
\end{widetext} 
and
\begin{align}
	c_\pm^\downarrow=\frac{(2 \alpha_\pm^\downarrow+1) \left(\gamma _1-\gamma _s\right)+\kappa- 2 \varepsilon }{2 \sqrt{3} \gamma _s}. \label{eqn:cD}
\end{align}
The spinor in Eq.~\eqref{eqn:WF_GeNWD} immediately gives the dispersion relation similar to the one for the upper block in Eq.~\eqref{eqn:dispRelGeU} 
\begin{align}
\frac{c_-^\downarrow}{c_+^\downarrow}=\frac{L^{m-2}_{\alpha_-^\downarrow+1}\left(\frac{R^2}{2}\right)
	L^{m}_{\alpha_+^\downarrow}\left(\frac{R^2}{2}\right)}{L^{m-2}_{\alpha_+^\downarrow+2}\left(\frac{R^2}{2}\right)
	L^{m}_{\alpha_-^\downarrow}\left(\frac{R^2}{2}\right)}. \label{eqn:dispRelGeD}
\end{align}
The implicit Eqs.~\eqref{eqn:dispRelGeU} and~\eqref{eqn:dispRelGeD} are solved numerically, yielding the energies $\varepsilon_m^s(n)$. Note that the eigenstates depend on $n$  via $\alpha$.

\subsection{Orbital corrections of the $g$ factor \label{sec:orbitalCorrections}}

\begin{figure*}[htb]
	\includegraphics{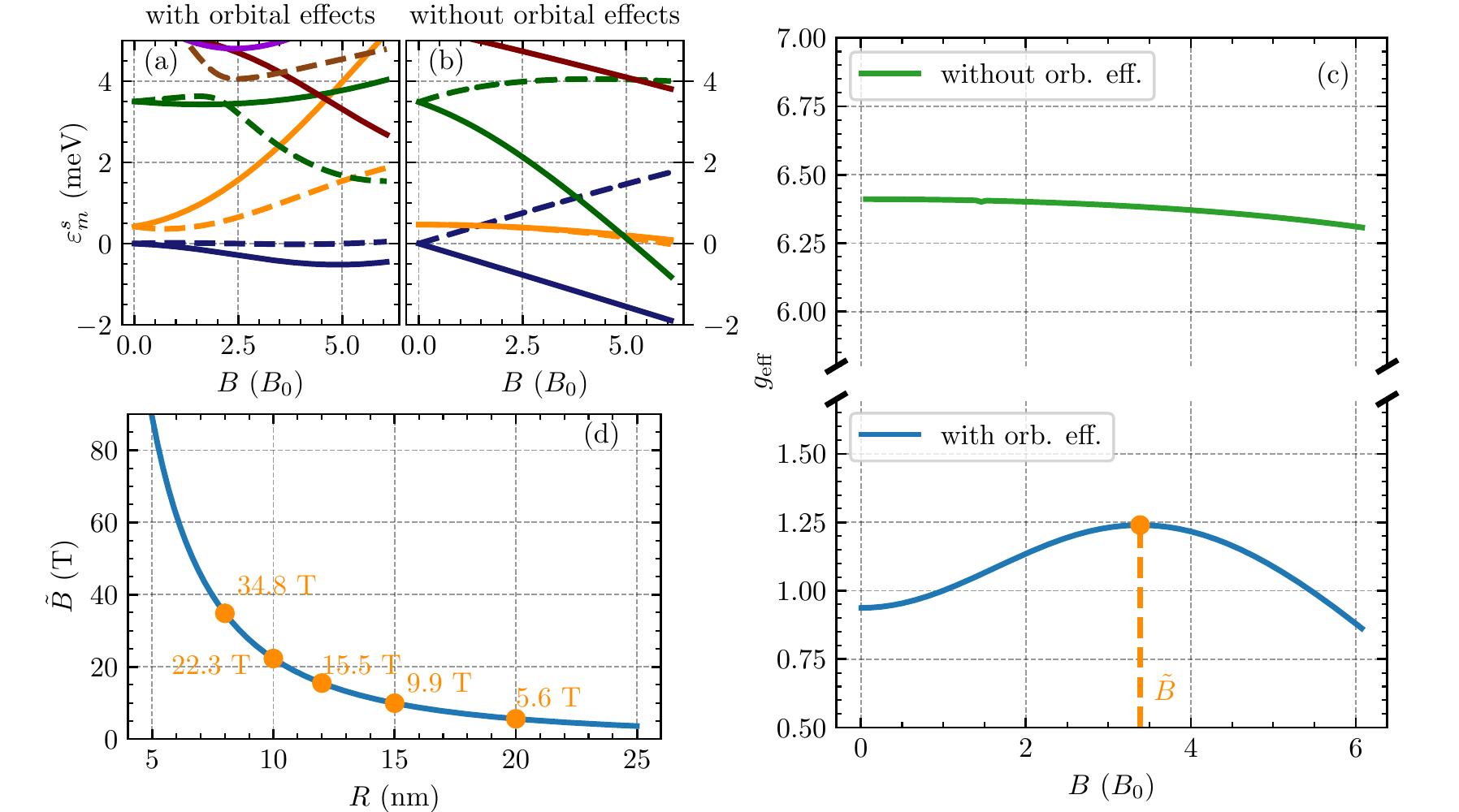}
	\caption{[(a),(b)] Energy spectrum of a Ge NW with circular cross section of radius $R=\SI{20}{\nano\meter}$ as a function of $B$ in units of the radius-dependent magnetic field $B_0$ at $k_z=0$. The colors indicate the Kramers partners. The solid (dashed) lines correspond to pseudospin $\uparrow(\downarrow)$ states. At $B=0$ we observe Kramers degeneracy that is lifted by a finite magnetic field. 
		(a) Energy spectrum calculated semi-analytically by numerically solving Eqs.~\eqref{eqn:dispRelGeU} and~\eqref{eqn:dispRelGeD}, in which we include orbital effects. 
		(b) Energy spectrum of the Hamiltonian $H$ defined in Eq.~\eqref{eqn:FullHamiltonian} calculated by numerically diagonalizing the discretized version of the Hamiltonian in Eq.~\eqref{eqn:FullHamiltonian} neglecting orbital effects (lattice spacing \SI{0.5}{nm}). The $g$ factor is strongly renormalized due to orbital effects. 
		(c) The effective $g$ factor $g_\mathrm{eff}$ [see Eq.~\eqref{eqn:effgFac}] obtained numerically using the lowest two states (blue lines) of panels (a) and (b). The green line depicts the effective $g$ factor without taking orbital effects into account, while the blue line does take them into account. The latter has a maximum at $\tilde{B}=3.4\,  B_0$. The effective $g$ factor only depends on the ratio $R^2/l_B^2$ and is seen to be strongly reduced by orbital magnetic field effects. 
		(d) The position $\tilde{B} = 3.4\, B_0$ of the maximum of the effective $g$ factor from the panel (c) as a function of the radius $R$. For smaller radius the maximum is located at stronger magnetic field. The functional dependence $\tilde{B}\propto 1/R^2$ is indicated by the blue line.\label{fig:gFacRen}
	}
\end{figure*}

The analytical solution of $H_{xy}$ [see Eq.~\eqref{eqn:HamPerp}] obtained in the previous section includes  exactly  orbital and Zeeman effects and, thus, it allows us to directly derive the $g$ factor of the NW. We demonstrate that the orbital magnetic fields lead to a correction of the $g$ factor by $\SI{400}{\percent}$ and discuss the reason for this strong renormalization. Here, we focus on a rather large NW of radius $R=\SI{20}{\nano\meter}$ where orbital effects are more pronounced.

Numerically solving Eq.~\eqref{eqn:dispRelGeU} and Eq.~\eqref{eqn:dispRelGeD}, we directly calculate the $g$ factor of the NW at $k_z=0$. In Fig.~\ref{fig:gFacRen}(a) we present the energy spectrum of the NW as a function of the magnetic field given in units of the radius-dependent magnetic field $B_0 = \hbar/e R^2 = \Phi_0/\pi R^2 =  \SI{658.2}{\tesla\times\nano\meter\squared\per R \squared}$, which is one flux quantum $\Phi_0=h/e$ through the cross section. We note that  $B/B_0=R^2/l_B^2$. By considering a  NW of radius $R=\SI{20}{\nano\meter}$, we obtain $B_0= \SI{1.65}{\tesla}$. 

For comparison we provide in Fig.~\ref{fig:gFacRen}(b) the spectrum for the same parameters without taking orbital effects into account ($\vect{A}=0$). We calculate the spectrum by numerically diagonalizing the discretized version of the Hamiltonian in Eq.~\eqref{eqn:FullHamiltonian}. In both plots, at weak magnetic fields, there is a large separation between the energies of the lowest two pairs of Kramers partners (orange and blue lines) and the states further above. Within the range up to $2\, B_0$ the ground state is close to the first excited state. However, only when orbital effects are taken into account, the first two pairs of former Kramers partners are coupled, resulting in an avoided crossing, which causes the orbital magnetic field induced reduction of the ground-state $g$ factor.

From the energies in Figs.~\ref{fig:gFacRen}(a) and~\ref{fig:gFacRen}(b) we directly deduce the effective $g$ factor as the difference between the $\downarrow$ and $\uparrow$ energies of the ground state Kramers partners 
\begin{align}
	g_\mathrm{eff} = \frac{\varepsilon^\downarrow_0- \varepsilon^\uparrow_2}{\mu_B B}. \label{eqn:effgFac}
\end{align}
Note that the QN $m$ is different for the $\uparrow$ and $\downarrow $ ground states due to the operators $a^2$ and $(a^\dagger)^2$ in $H_{xy}$ [see Eq.~\eqref{eqn:HamPerp}].
In Fig.~\ref{fig:gFacRen}(c), we show the down-renormalization of the NW $g$ factor by $\SI{400}{\percent}$ due to orbital effects.
This large renormalization can be understood from the spectra shown in Figs.~\ref{fig:gFacRen}(a) and~\ref{fig:gFacRen}(b). At $B=0$, the ground state Kramers pair (blue lines) is close to the first excited pairs (orange lines). At finite $B$, these pairs are coupled by orbital effects, yielding an avoided crossing [cf. Fig.~\ref{fig:gFacRen}(a)] that strongly reduces the $g$ factor.
Note that for the calculation of the $g$ factor without orbital effects we always take the difference between the dashed and solid blue line and ignore the crossing with higher energy Kramers partners (orange and green lines).

Interestingly, we observe a maximum of the effective $g$ factor with orbital effects of $g_\mathrm{eff} = 1.24$ at the magnetic field $B=\tilde{B} = 3.4\, B_0$, which decreases at larger values of $R$. This trend is illustrated in Fig.~\ref{fig:gFacRen}(d), where we show that $\tilde{B}\propto 1/R^2$. We also provide the values of $\tilde{B}$ for some specific radii and we conclude that the maximum of the $g$ factor can be reached at realistic values of magnetic fields only in rather wide NWs.

\subsection{Effective Hamiltonian} \label{sec:EffHam}
We use the exact solution of the Hamiltonian at $k_z = 0$ to construct a simple effective low-energy theory that models the dynamics of holes in NWs and long QDs. We  treat the $k_z$-dependent terms, $H_{zz}$  and $H_\mathrm{int}$, of the full ILK Hamiltonian $H_{\rm ILK}$ in Eq.~\eqref{eqn:Factorization} as perturbation. Applying the definition of the Landau level operators $a$ from Eq.~\eqref{eqn:LandauLevelOp}, the interaction part $H_\mathrm{int}$ can be rewritten in the spin basis ($+3/2$, $-1/2$, $-3/2$, $+1/2$) as
\begin{align}
	\frac{H_\mathrm{int}}{\hbar \omega_c} &= \sqrt{6} \gamma_s 
	\begin{pmatrix}
		0 & 0 & 0 & -a \\
		0 & 0 & a & 0\\
		0 & a^\dagger & 0 & 0\\
		-a^\dagger & 0 & 0 & 0
	\end{pmatrix} l_B.
\end{align}
Using  Eqs.~\eqref{eqn:WF_GeNWU} and~\eqref{eqn:WF_GeNWD}, we calculate the eigenfunctions of $H_{xy}$ [see Eq.~\eqref{eqn:HamPerp}] which we arrange in the four-dimensional spinors in the same spin basis 
\begin{align}
	\varphi^\uparrow_m &= \left(\Psi_{m}^{+3/2}, \Psi_{m}^{-1/2},0, 0\right),\\
	\varphi^\downarrow_m &= \left(0, 0, \Psi_{m}^{-3/2}, \Psi_{m}^{+1/2}\right).
\end{align}
In order to model the effects of external electric fields induced by the metallic gates, we consider a multipole expansion of the electrostatic potential. In particular, we consider a homogeneous electric field $\vect{E} = (E_x, E_y, 0)$ and a quadratic potential parametrized by a matrix $\delta E_{ij}$. Explicitly, we consider the electrostatic energy in the form
\begin{align}
&	H_E = e E_x r \cos \varphi +e E_y r \sin\varphi \label{eqn:Heff_efield}\\
	& +\frac{e}{2} (\delta E_{xx} \cos^2 \varphi + \delta E_{yy} \sin^2 \varphi + \delta E_{xy} \cos\varphi\sin\varphi)  r^2. \nonumber 
\end{align}
This potential accurately describes the electrostatic potential in several experimental setups comprising multiple gates~\cite{Bosco2021b}. As we are studying an isotropic NW with circular cross section, without loss of generality, we now fix the direction of the homogeneous field to the $x$ direction ($E_y = 0$). 
The inhomogeneous term can be neglected in narrow NWs, while in thick NWs the inhomogeneous electric field becomes relevant and we include it in our analysis.

The low-energy holes are well described by an effective Hamiltonian comprising the three lowest Kramers partners. Explicitly, this Hilbert space is spanned by the six component basis $(\varphi^\uparrow_2, \varphi^\uparrow_1, \varphi^\uparrow_0, \varphi^\downarrow_0, \varphi^\downarrow_1, \varphi^\downarrow_2)$ consisting of the lowest six states in Fig.~\ref{fig:gFacRen}(a) (at $B=0$). The states correspond to the solid-blue, solid-orange, solid-green, dashed-blue, dashed-orange, and dashed-green lines in Fig.~\ref{fig:gFacRen} in the order as they appear in the basis. For all these states the QN  is $n=1$ and if not stated otherwise we assume $n=1$ in the following. States with larger $n$ lie at higher energies and, thus, are neglected here. In this basis the effective Hamiltonian $H_\mathrm{LK}^\mathrm{eff}$ can be expressed as
\begin{widetext}
	\begin{small}
			\begin{align}
			H_\mathrm{LK}^\mathrm{eff}= \left(
			\begin{array}{cccccc}
				\frac{\hbar^2 k_z^2}{2 m_2^\uparrow}+\varepsilon _2^\uparrow  +q_2^\uparrow \delta E_+& i d_1^\uparrow E_x & q^\uparrow \delta E_- & 0 & \alpha _1 k_z & 0 \\
				-i d_1^\uparrow E_x & \frac{\hbar^2 k_z^2}{2 m_1^\uparrow}+\varepsilon _1^\uparrow  +q_1^\uparrow \delta E_+ & i d_2^\uparrow E_x & \alpha _2 k_z  & 0 & 0 \\
				q^\uparrow\delta E_-^\ast & -i d_2^\uparrow E_x & \frac{\hbar^2 k_z^2}{2 m_0^\uparrow}+\varepsilon _0^\uparrow  +q_0^\uparrow \delta E_+ & 0 & 0 & 0 \\
				0 &  \alpha _2 k_z & 0 & \frac{\hbar^2 k_z^2}{2 m_0^\downarrow}+\varepsilon _0^\downarrow  +q_0^\downarrow \delta E_+& i d_1^\downarrow E_x & q^\downarrow \delta E_- \\
				\alpha _1 k_z & 0 & 0 & -i d_1^\downarrow E_x & \frac{\hbar^2 k_z^2}{2 m_1^\downarrow}+\varepsilon _1^\downarrow   +q_1^\downarrow \delta E_+& i d_2^\downarrow E_x \\
				0 & 0 & 0 & q^\downarrow\delta E_-^\ast & -i d_2^\downarrow E_x & \frac{\hbar^2 k_z^2}{2 m_2^\downarrow}+\varepsilon _2^\downarrow  +q_2^\downarrow \delta E_+\\
			\end{array}
			\right), \label{eqn:effHam}
		\end{align}
	\end{small}
\end{widetext}
where we defined
\begin{align}
	\delta E_+ &= \delta E_{xx} +\delta E_{yy},\\
	\delta E_- &= \delta E_{xx} -\delta E_{yy} - i \delta E_{xy}.
\end{align}
We note that the parameters $d_{1,2}^{\uparrow,\downarrow}$, $q^{\uparrow,\downarrow}$, $q_{0, 1, 2}^{\uparrow,\downarrow}$, and $\alpha_{1,2}$ are real and will be defined explicitly below.
The energies  
\begin{align}
	\varepsilon_m^s &= \mel{\varphi_m^s}{H_{xy} + H_Z + V}{\varphi_m^s} 
\end{align}
($s= \uparrow, \downarrow$) are obtained numerically from the implicit relations in Eqs.~\eqref{eqn:dispRelGeU} and \eqref{eqn:dispRelGeD}. The expressions for effective masses $m_i^s$ ($i=0,1,2$) consist of two contributions. The first one arises from $H_{zz}$ as $\mel{\varphi_i^s}{H_{zz}}{\varphi_j^s} $, which simplifies to the matrix elements with $i=j$, since in our case there are only diagonal matrix entries. The second contribution of order $k_z^2$ arises from $H_\mathrm{int}$ in second-order perturbation theory and is also diagonal. The total effective mass is then given by~\cite{Bosco2021b}
\begin{align}
	\frac{\hbar^2}{2 m_i^s} &= \mel{\varphi_i^s}{H_{zz}}{\varphi_i^s} +\sum\limits_{l, n} \frac{\left|\mel{\varphi_i^s}{H_\mathrm{int}}{\varphi_l^s(n)}\right|^2}{\varepsilon^s_i - \varepsilon^s_l(n)}.\label{eqn:GEmeff}
\end{align}
The sum runs over all states outside the considered subspace.
 We observe strong couplings $\mel{\varphi_i^s}{H_\mathrm{int}}{\varphi_l^s(n)}$ to states with large $n$ that lead to considerable perturbative contributions to the effective mass. Thus, to ensure the convergence of the perturbation theory numerically, we take into account states up to $n=10$. 

Next, we calculate the SOI terms that couple the $\uparrow$ and $\downarrow$ blocks as the following overlaps between HH and LH states:
\small
\begin{align}
	\alpha_1 &= \sqrt{6} \gamma_s \hbar\omega_c l_B \left(\mel{\Psi_2^{-1/2}}{a}{\Psi_1^{-3/2}} -\mel{\Psi_2^{+3/2}}{a}{\Psi_1^{+1/2}}\right), \label{eqn:SOcouplingGe1}\\
	\alpha_2 &= \sqrt{6} \gamma_s \hbar\omega_c l_B \left(\mel{\Psi_1^{-1/2}}{a}{\Psi_0^{-3/2}}-\mel{\Psi_1^{+3/2}}{a}{\Psi_0^{+1/2}}\right). \label{eqn:SOcouplingGe2}
\end{align}\normalsize
The electric dipole moments, which result from the first term of Eq.~\eqref{eqn:Heff_efield}, are given by 
\begin{align}
	d_1^s &= \frac{e}{2 i} \left(\mel{\Psi_2^{\pm 3/2}}{r}{\Psi_1^{\pm 3/2}} + \mel{\Psi_2^{\mp 1/2}}{r}{\Psi_1^{\mp 1/2}}\right), \label{eqn:dipCoup1}\\
	d_2^s &= \frac{e}{2 i} \left(\mel{\Psi_1^{\pm 3/2}}{r}{\Psi_0^{\pm 3/2}} + \mel{\Psi_1^{\mp 1/2}}{r}{\Psi_0^{\mp 1/2}}\right), \label{eqn:dipCoup2}
\end{align}
and the quadrupole moments from the second term are given by
\begin{align}
	q^s &= \frac{e}{8} \left(\mel{\Psi_2^{\pm 3/2}}{r^2}{\Psi_0^{\pm 3/2}} + \mel{\Psi_2^{\mp 1/2}}{r^2}{\Psi_0^{\mp 1/2}}\right) \label{eqn:quadCoup1}
\end{align}
and
\begin{align}
	q_i^s = \frac{e}{4}\left(\mel{\Psi_i^{\pm 3/2}}{r^2}{\Psi_i^{\pm 3/2}} + \mel{\Psi_i^{\mp 1/2}}{r^2}{\Psi_i^{\mp 1/2}}\right) \label{eqn:quadCoup2}
\end{align}	
($i = 0, 1, 2$).

For a discussion of the behavior of  the effective parameters as a function of the magnetic field we refer to Appendix~\ref{sec:effParam}. In the next section, we construct an effective low-energy theory of the two lowest energy states by integrating out the states at higher energy.

\subsection{$2\times 2$ Wire Hamiltonian \label{sec:2x2wireHam}}

\begin{figure*}[htb]
	\includegraphics{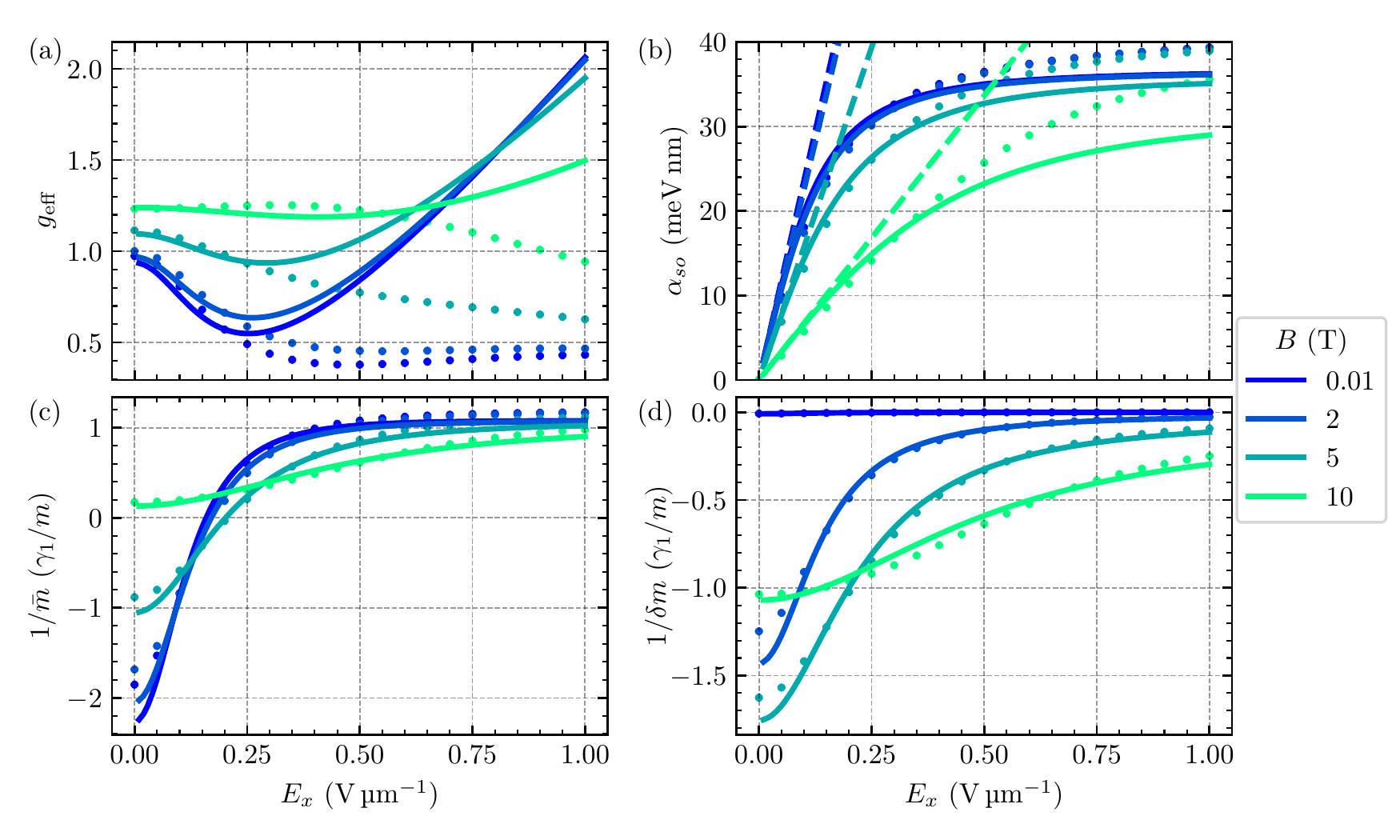}
	\caption{Effective parameters of the effective $2\times 2$ model $H^{2\times 2}$ (solid lines) as a function of the homogeneous electric field $E_x$ for a Ge NW with radius $R=\SI{15}{\nano\meter}$. The dots show the same parameters calculated numerically with the Hamiltonian in Eq.~\eqref{eqn:FullHamiltonian} discretized in real space (lattice spacing \SI{0.25}{\nano\meter}).
		(a) Effective $g$ factor $g_\mathrm{eff}$ according to Eq.~\eqref{eqn:effgfac}. At weak $B$, the results from perturbation theory are valid only at rather weak electric fields ($E_x<\SI{0.25}{\volt\per\micro\meter}$) while at strong $B$, the perturbation theory can be extended to $E_x<\SI{0.5}{\volt\per\micro\meter}$. At stronger electric field the perturbative $g$ factor increases unphysically,  while it should decrease and stay at a small value as the numerical results show.
		(b) The effective SOI coefficient $\alpha_{so}$ [cf. Eq.~\eqref{eqn:effSOI}] exhibits a linear behavior at weak electric field as provided by the expansion of Eq.~\eqref{eqn:effSOIlin}. At stronger electric field $\alpha_{so}$ saturates and with increasing $B$ it decreases due to orbital effects. 
		(c) Inverse average effective mass $1/\bar{m}$ from Eq.~\eqref{eqn:effMassAv} and 
		(d) inverse spin-dependent mass $1/\delta m$ from Eq.~\eqref{eqn:effMassDiff}. The average mass can take negative values at weak electric field and approaches the average HH-LH mass $m/\gamma_1$ at stronger $E_x$. The spin-dependent mass is zero at $B=0$ due to time-reversal symmetry and decreases with $E_x$ at $B \neq 0$.
	The electric field dependence of the mass terms and the SOI is captured well by the perturbation theory up to $E_x = \SI{1}{\volt\per\micro\meter}$. However, $\alpha_{so}$ is underestimated for strong electric fields. The mass terms from numerical calculations deviate visibly from the perturbative result at $E_x=0$ due to the limited number of states that can be taken into account for the calculation of the term coming from $H_{int}$ in Eq.~\eqref{eqn:GEmeff} in the numerics. \label{fig:Heff_SOImeffGe_R15}}
\end{figure*}

To obtain a simple NW Hamiltonian describing the two states lowest in energy, we start with the Hamiltonian $ 	H_\mathrm{LK}^\mathrm{eff}$ introduced in Eq.~\eqref{eqn:effHam}, resort to a second order perturbation theory, and project onto the low-energy $2\times 2$ subspace. This procedure is well justified when the energy scale characterizing the holes in the $z$ direction is much smaller than the subband gap $\varepsilon_0^\downarrow - \varepsilon_1^\downarrow$. Without strain and at weak $B$ field, this condition puts a strong constraint on the possible confinement energy in the $z$ direction because the subband energy gap is rather small, cf. Fig.~\ref{fig:gFacRen}(a). Larger values of the confinement along the NW can be achieved  at stronger $B$ fields or by considering strain, as we discuss in Sec.~\ref{sec:strain}. 
We now focus on  the effect of the electrostatic potential.

\subsubsection{Homogeneous electric field limit}
We first consider a homogeneous electric field $E_x$. Since we are interested in the ground-state $(\varphi^\uparrow_2, \varphi^\downarrow_0)$, we can restrict ourselves to the analysis of the ground state pair and the pair of states $(\varphi^\uparrow_1, \varphi^\downarrow_1)$ coupled to it via the SOI $\alpha_{1,2}$ and the dipole coupling $d^s_{1}$. The states $\varphi_0^\uparrow$ and $\varphi_2^\downarrow$ are decoupled from the ground state, and thus we assume that they have only a weak influence. We also introduce the following averages and differences of energies
\begin{align}
	\varepsilon^\uparrow_\pm &= \frac{\varepsilon _1^\uparrow \pm\varepsilon _2^\uparrow }{2}, \label{eqn:enDiffU}\\
	\varepsilon^\downarrow_\pm &= \frac{\varepsilon _1^\downarrow \pm\varepsilon _0^\downarrow }{2}. \label{eqn:enDiffD}
\end{align}
Additionally, we define the angles
\begin{align}
	\tan\theta^s &= \frac{\varepsilon_-^s + \Omega^s}{d_1^s E_x} \label{eqn:orbAng}
\end{align}
with $s=\uparrow,\downarrow$ and with the energies
\begin{align}
	\Omega^s &= \sqrt{\left(d_1^s E_x\right)^2 + \left(\varepsilon_-^s \right)^2} \label{eqn:orbEn}
\end{align}
being dependent on the electric field $E_x$.
Next, we rotate the $\uparrow$ and $\downarrow$ blocks by $\theta^\uparrow$ and $\theta^\downarrow$, respectively, as described in Appendix~\ref{sec:effMod}. Working  with second-order perturbation theory and projecting the results onto the ground-state subspace $(\varphi^\uparrow_2, \varphi^\downarrow_0)$, we obtain up to second order in $k_z$ 
\begin{align}
	H^{2\times 2} = 
	\frac{\hbar^2}{2 \bar{m}} k_z^2 + \frac{1}{2}\left(g_\mathrm{eff} \mu_B B+\frac{\hbar^2 k_z^2}{\delta m} \right)\sigma_z - \alpha_{so} k_z \sigma_y . \label{eqn:Heff2x2}
\end{align}
In some cases a cubic SOI term becomes relevant, which requires to extend the perturbation theory to third order~\cite{Hetenyi2022}.
The effective $g$ factor is electric field dependent and is given by 
\begin{align}
	\mu_B B g_\mathrm{eff} = \varepsilon_+^\downarrow -\varepsilon_+^\uparrow -\Omega^\downarrow+\Omega^\uparrow.\label{eqn:effgfac}
\end{align}
Moreover, the direct Rashba SOI~\cite{Kloeffel2011,Kloeffel2018} is given by
\begin{align}
	\alpha_{so} = &\alpha_1  \cos(\theta^\downarrow) \sin(\theta^\uparrow) - \alpha_2 \cos(\theta^\uparrow) \sin(\theta^\downarrow) \label{eqn:effSOI}\\
	= &\frac{ d_1^\downarrow\alpha_1 \varepsilon_-^\uparrow - d_1^\uparrow\alpha_2 \varepsilon_-^\downarrow}{\varepsilon_-^\downarrow \varepsilon_-^\uparrow} E_x + \mathcal{O}(E_x^3),\label{eqn:effSOIlin}
\end{align}
and it is linear at weak electric field. We also introduce the average effective mass
\begin{align}
&	\frac{1}{\bar{m}} = \frac{m_1^\uparrow + m_2^\uparrow - (m_1^\uparrow - m_2^\uparrow)\cos(2 \theta^\uparrow)}{ 4 m_1^\uparrow m_2^\uparrow}\nonumber\\
	&+\frac{m_0^\downarrow + m_1^\downarrow + (m_0^\downarrow - m_1^\downarrow)\cos(2 \theta^\downarrow)}{ 4 m_0^\downarrow m_1^\downarrow}\nonumber\\
	&-\frac{m}{\hbar \omega_c l_B^2}\Bigg( \frac{\left[\alpha_1 \sin(\theta^\uparrow) \sin(\theta^\downarrow) + \alpha_2  \cos(\theta^\uparrow) \cos(\theta^\downarrow)\right]^2}{\varepsilon_+^\downarrow - \varepsilon_+^\uparrow + \Omega^\uparrow + \Omega^\downarrow}\nonumber\\
	&+ \frac{\left[\alpha_1 \cos(\theta^\uparrow) \cos(\theta^\downarrow) + \alpha_2  \sin(\theta^\uparrow) \sin(\theta^\downarrow)\right]^2}{\varepsilon_+^\uparrow - \varepsilon_+^\downarrow + \Omega^\uparrow + \Omega^\downarrow} \Bigg), \label{eqn:effMassAv}
\end{align}
and the spin-dependent mass
\begin{align}
	&\frac{1}{\delta m} = \frac{m_1^\uparrow + m_2^\uparrow - (m_1^\uparrow - m_2^\uparrow)\cos(2 \theta^\uparrow)}{ 4 m_1^\uparrow m_2^\uparrow}\nonumber\\
	&-\frac{m_0^\downarrow + m_1^\downarrow + (m_0^\downarrow - m_1^\downarrow)\cos(2 \theta^\downarrow)}{ 4 m_0^\downarrow m_1^\downarrow}\nonumber\\
	&+ \frac{m}{\hbar\omega_c l_B^2}\Bigg(\frac{\left[\alpha_1 \cos(\theta^\uparrow) \cos(\theta^\downarrow) + \alpha_2  \sin(\theta^\uparrow) \sin(\theta^\downarrow)\right]^2}{\varepsilon_+^\uparrow - \varepsilon_+^\downarrow + \Omega^\uparrow + \Omega^\downarrow}\nonumber\\
	&- \frac{\left[\alpha_1 \sin(\theta^\uparrow) \sin(\theta^\downarrow) + \alpha_2  \cos(\theta^\uparrow) \cos(\theta^\downarrow)\right]^2}{\varepsilon_+^\downarrow - \varepsilon_+^\uparrow + \Omega^\uparrow + \Omega^\downarrow}\Bigg). \label{eqn:effMassDiff}
\end{align}
We note that both masses $\delta m$ and $\bar{m}$ inherit an electric field dependence by the angles $\theta^s$ and energies $\Omega^s$.

In Fig.~\ref{fig:Heff_SOImeffGe_R15}(a), we study the effective $g$ factor $g_\mathrm{eff}$ as a function of the electric field $E_x$ for different values of the magnetic field $B$ (solid lines). For a comparison we provide results from numerical calculations (dots) where we diagonalize the discretized version of the Hamiltonian in Eq.~\eqref{eqn:FullHamiltonian} including the homogeneous electric field given by Eq.~\eqref{eqn:Heff_efield}. The comparison shows that the perturbation theory gives a good estimate for $g_\mathrm{eff}$ at weak electric field. At $E_x\approx\SI{0.25}{\volt\per\micro\meter}$ ($E_x\approx\SI{0.5}{\volt\per\micro\meter}$ and $B\geq\SI{5}{\tesla}$) the perturbation theory starts to fail and predicts an unphysical increase of $g_\mathrm{eff}$.
The effective SOI [cf. Fig.~\ref{fig:Heff_SOImeffGe_R15}(b)] $\alpha_{so}$ increases linearly at weak $E_x$ according to Eq.~\eqref{eqn:effSOIlin} [illustrated by the dashed lines in Fig.~\ref{fig:Heff_SOImeffGe_R15}(b)], and it saturates at stronger electric field ($E_x\gtrsim \SI{0.5}{\volt\per\micro\meter}$). Due to orbital effects, we observe a decrease of $\alpha_{so}$ with increasing magnetic field. For the realization of MBSs in the Ge NW, a weak electric field is favorable because there the effective $g$ factor is not so strongly suppressed, enabling to reach the topological phase~\cite{Lutchyn2010,Oreg2010,Alicea2010,Maier2014} at lower magnetic fields,  away from the critical magnetic field of the superconductor. We note that, if the NW is in proximity to the thin bulk superconductor, there will be an additional renormalization of NW parameters, so-called metallization~\cite{Reeg2017,Reeg2018, Reeg2018a,Woods2019,Winkler2019,Kiendl2019}, which needs to be taken into account.

In addition, in Figs.~\ref{fig:Heff_SOImeffGe_R15}(c) and~\ref{fig:Heff_SOImeffGe_R15}(d), we present the inverse average effective mass $1/\bar{m}$ and spin-dependent mass $1/\delta m$, respectively. Interestingly, the inverse average mass is negative at weak electric field $E_x\lesssim \SI{0.2}{\volt\per\micro\meter}$ and $B\lesssim \SI{9}{\tesla}$. It approaches the average HH-LH mass $\gamma_1/m$ at stronger electric fields. As expected from time-reversal symmetry at $B = 0$, the inverse spin-dependent mass term $1/\delta m=0$~\cite{Adelsberger2022}. It is relevant only at weak electric fields and we observe that it vanishes at strong $E_x$.
We find a simple formula for the inverse effective masses in the limit $E_x\rightarrow 0$,
\begin{align}
	\lim\limits_{E_x\rightarrow0} \bar{m}^{-1} &= \frac{m_0^\downarrow + m_2^\uparrow}{2 m_0^\downarrow m_2^\uparrow} - \frac{m}{\hbar\omega_c l_B^2}\Bigg( \frac{2\alpha_1^2}{\varepsilon_+^\uparrow-\varepsilon_+^\downarrow+\abs{\varepsilon_-^\uparrow}+\abs{\varepsilon_-^\downarrow}}\nonumber\\
	& + \frac{2\alpha_2^2}{-\varepsilon_+^\uparrow+\varepsilon_+^\downarrow+\abs{\varepsilon_-^\uparrow}+\abs{\varepsilon_-^\downarrow}}\Bigg), \\
	\lim\limits_{E_x\rightarrow0} \delta m^{-1} &=  \frac{m_0^\downarrow - m_2^\uparrow}{2 m_0^\downarrow m_2^\uparrow} + \frac{m}{\hbar\omega_c l_B^2}\Bigg(
	\frac{2\alpha_2^2}{-\varepsilon_+^\uparrow+\varepsilon_+^\downarrow+\abs{\varepsilon_-^\uparrow}+\abs{\varepsilon_-^\downarrow}}\nonumber\\ &- \frac{2\alpha_1^2}{\varepsilon_+^\uparrow-\varepsilon_+^\downarrow+\abs{\varepsilon_-^\uparrow}+\abs{\varepsilon_-^\downarrow}}\Bigg).
\end{align}
These formulas make manifest that the spin-dependent mass is dominated by the average mass of the states $\varphi_0^\downarrow$ and  $\varphi_2^\uparrow$ $(m_0^\downarrow\pm m_2^\uparrow)/2 m_0^\downarrow m_2^\uparrow$, but it acquires a correction by the SOI coefficients $\alpha_{1,2}$. The results for the SOI and the mass terms from perturbation theory agree well with the numerical results in the displayed range of electric field; the SOI is underestimated at strong $E_x$ by the perturbation theory.

\begin{figure*}[!t]
	\includegraphics{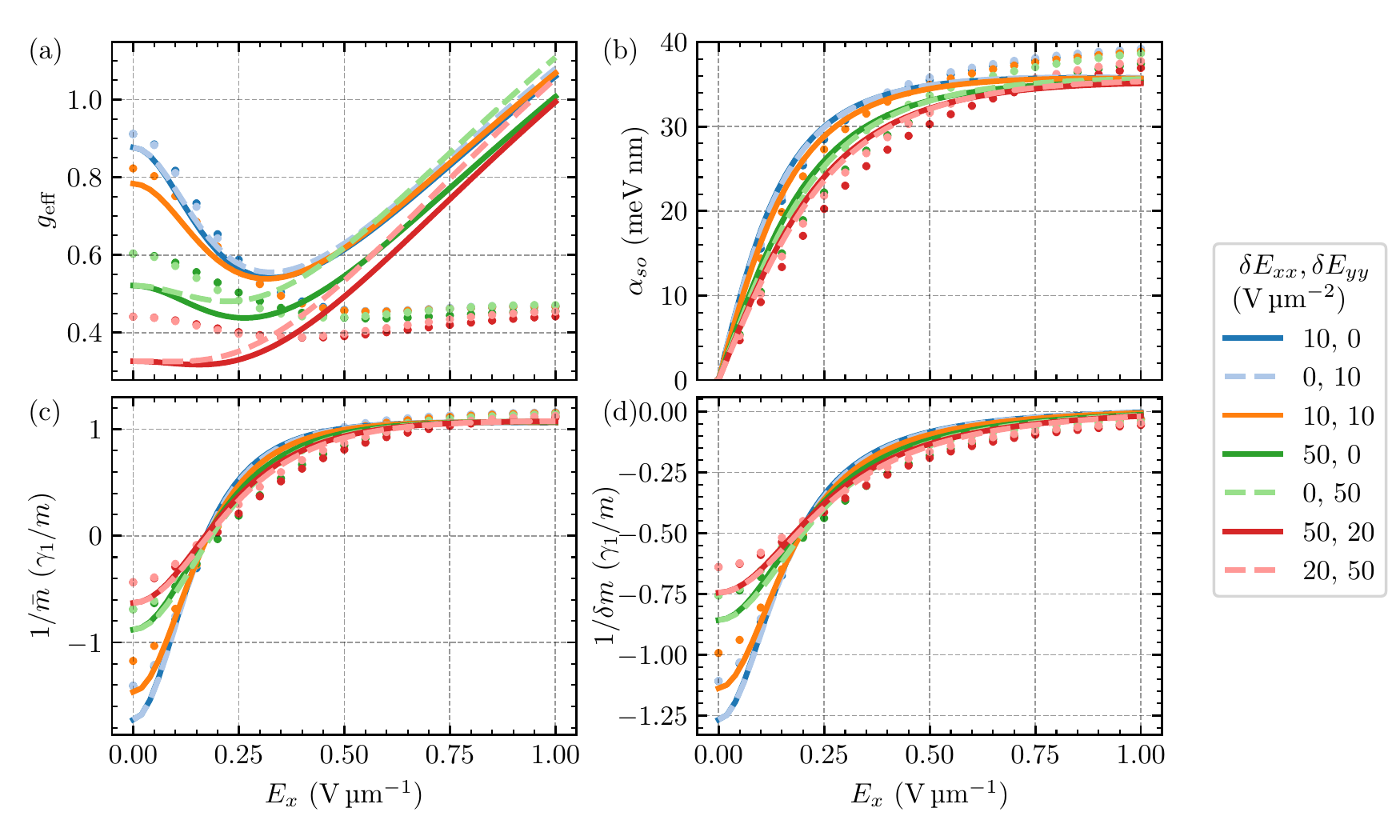}
	\caption{Effective parameters of the  $2\times 2$ model described by $H^{2\times 2}$ in Eq.~\eqref{eqn:Heff2x2} for a circular Ge NW are shown here, including an inhomogeneous electric field $\delta E_{xx}$ and $\delta E_{yy}$ calculated perturbatively (solid and dashed lines) and numerically by diagonalizing the Hamiltonian in Eq.~\eqref{eqn:FullHamiltonian} including the inhomogeneous electric field term in Eq.~\eqref{eqn:Heff_efield} discretized in real space (dots) at $B=\SI{2}{\tesla}$ as a function of the homogeneous electric field $E_x$. Here, we assume the radius $R=\SI{15}{\nano\meter}$ for the cross section of the NW and a lattice spacing of $\SI{0.25}{\nano\meter}$ for the numerical calculations. In general the effect of the electric field gradient is pronounced at weak $E_x$. At $E_x = \SI{1}{\volt\per\micro\meter}$ the effective parameters are renormalized only slightly by the inhomogeneous field. The main reason for the renormalization of the parameters are the diagonal quadrupole moment terms $q_i^s$ ($i=0,1,2$) and thus we observe a strong effect at $\delta E_{xx} = \delta E_{yy} = \SI{10}{\volt\per\micro\meter\squared}$ where the off-diagonal terms in the Hamiltonian in Eq.~\eqref{eqn:effHam} vanish.
		(a) The effective $g$ factor $g_\mathrm{eff}$ is reduced by the inhomogeneous field at weak $E_x$. The effective model overestimates the effect of the inhomogeneous field at weak $E_x$ and becomes inaccurate at strong $E_x$ in agreement with Fig.~\ref{fig:Heff_SOImeffGe_R15}(a).
		(b) The effective SOI strength $\alpha_{so}$ decreases with increasing inhomogeneous field slightly. 
		(c) The inverse average effective mass $1/\bar{m}$ moves closer to zero at $E_x = 0$ and still approaches a value close to $\gamma_1/m$ at strong electric field. 
		(d) The inverse spin-dependent mass $\abs{1/\delta m}$ is reduced by the inhomogeneous electric field at $E_x=0$ and approaches zero at strong $E_x$.
		The SOI and the mass terms are well described by the effective model as the comparison with the numerical result shows.
		 \label{fig:Heff_SOImeffGe_R15_6x6_B2}}
\end{figure*}

\subsubsection{Inhomogeneous electric field}

In thick NWs, the approximation of a homogeneous electric field is not well-justified and there can be corrections arising from inhomogeneity of the electric field~\cite{Bosco2021b}, captured by the Hamiltonian $H_\mathrm{LK}^\mathrm{eff}$ defined in Eq.~\eqref{eqn:effHam}.
For simplicity we neglect the inhomogeneous electric field term $\delta E_{xy}$ in Eq.~\eqref{eqn:Heff_efield} since the final results for the effective parameters only depend on the absolute value of the total inhomogeneous electric field. We find that the quadrupole moment terms can strongly renormalize the effective parameters, in particular, the $g$ factor at weak $E_x$.

 In analogy to the homogeneous electric field limit, we derive an effective Hamiltonian $H^{2\times 2}$  for the lowest two states in second order perturbation theory. We arrive at the same form of the effective $2\times 2$ Hamiltonian as in Eq.~\eqref{eqn:Heff2x2}.
 However, the effective masses and the SOI depend now also  on the electric field gradient $\delta E_\pm$, see Fig.~\ref{fig:Heff_SOImeffGe_R15_6x6_B2}. 
 The electric field dependence of the effective parameters is calculated perturbatively starting from the analytical result at $E_x = \delta E_{xx} = \delta E_{yy} = 0$. We analyze the same electric field range as in Fig.~\ref{fig:Heff_SOImeffGe_R15} and focus on weak inhomogeneous electric fields, where the qualitative dependence on $E_x$ stays as for $\delta E_{xx} = \delta E_{yy} = 0$. The effect of the inhomogeneous electric field on the effective parameters displayed in Fig.~\ref{fig:Heff_SOImeffGe_R15_6x6_B2} is strong at weak homogeneous electric field. The main effect is coming from the diagonal quadrupole moment terms $q_i^s$ ($i=0,1,2$) which cause an enhancement of the subband gap between the lowest Kramers pair and the states higher in energy. This enhancement results in a significant renormalization of the $g$ factor at $\delta E_{xx} = \delta E_{yy} = \SI{10}{\volt\per\micro\meter\squared}$ where the off-diagonal quadrupole moment terms in  the Hamiltonian in Eq.~\eqref{eqn:effHam} vanish. The effective $g$ factor and the masses tend closer to zero with increasing inhomogeneous electric field at $E_x=0$. For the $g$ factor this effect is overestimated by the effective model while it is underestimated for the mass terms as the comparison to the numerical results (dots) shows. As expected from  Fig.~\ref{fig:Heff_SOImeffGe_R15}(a) the effective model fails to predict the electric field dependence of the effective $g$ factor at strong homogeneous electric field correctly. However, the effective model describes the SOI and the mass terms well.

\subsection{Strain \label{sec:strain}}

\begin{figure*}[htb]
	\includegraphics{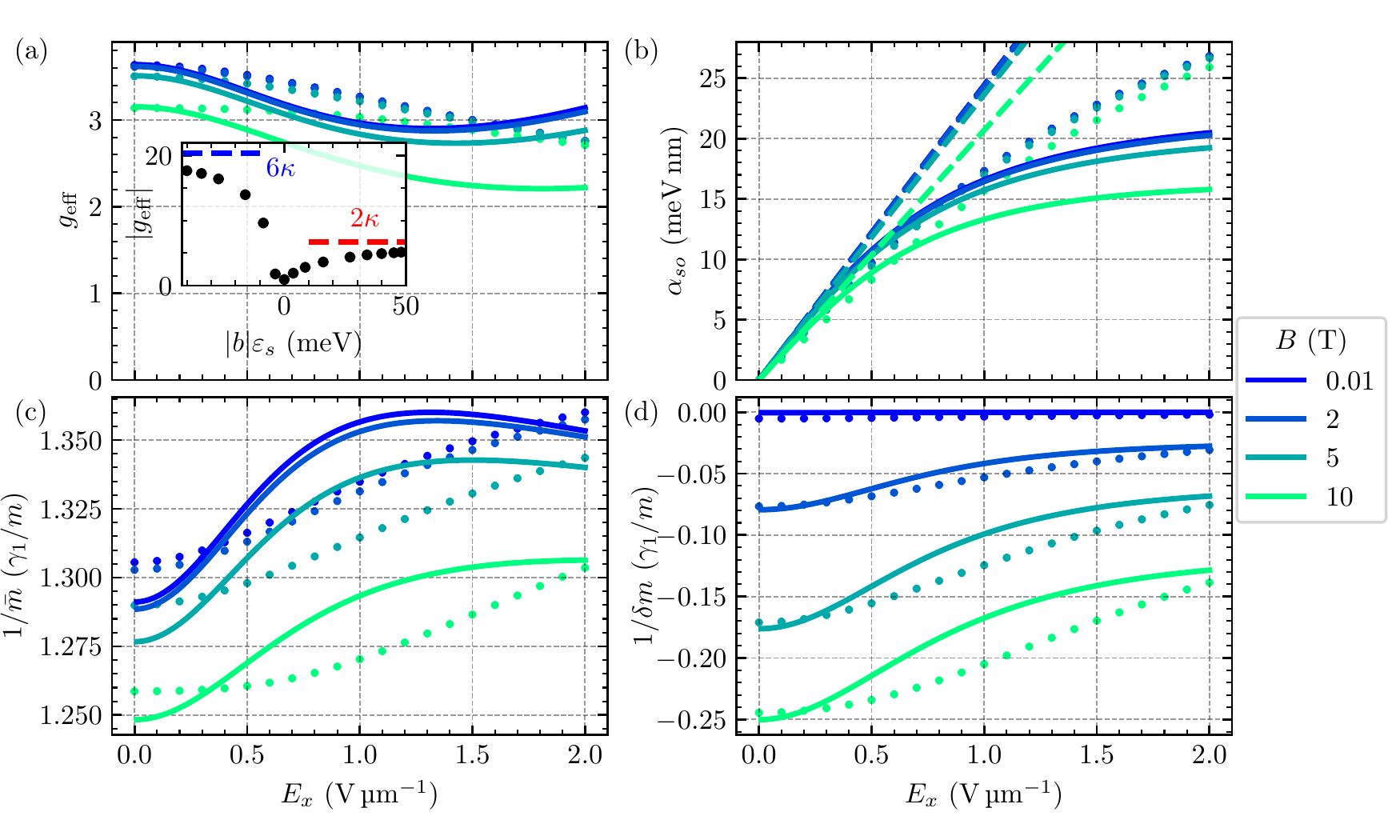}
	\caption{Effective parameters of the a strained core/shell NW of radius $R = \SI{15}{nm}$ according to the effective $2\times 2$ model, $H^{2\times 2}$ in Eq.~\eqref{eqn:Heff2x2} (solid lines), as a function of the electric field $E_x$ for strain energy $\abs{b} \varepsilon_s = \SI{15.7}{\milli\electronvolt}$ (corresponds to $\gamma = 0.1$). Strain is included via the BP Hamiltonian $	H_\textrm{BP} $ defined in Eq.~\eqref{eqn:BPHam}. The dots show the same quantities calculated numerically by diagonalizing the Hamiltonian in Eq.~\eqref{eqn:FullHamiltonian} discretized in real space (lattice spacing \SI{0.25}{\nano\meter})
		(a) The effective $g$ factor [see Eq.~\eqref{eqn:effgfac}] is large at weak $E_x$ and decreases throughout the whole depicted range. The perturbatively calculated $g$ factor deviates from this behavior at strong electric field and increases after reaching a minimum as in Fig.~\ref{fig:Heff_SOImeffGe_R15}(a). The inset in (a) shows $g_\mathrm{eff}$ calculated semi-analytically according to Eqs.~\eqref{eqn:dispRelGeU}, \eqref{eqn:dispRelGeD}, and~\eqref{eqn:effgFac} as a function of the strain energy $\abs{b} \varepsilon_s$ at $B=\SI{0.01}{\tesla}$. The result is almost independent of the strength of the magnetic field. Strain increases $g_\mathrm{eff}$ independently of the sign of $\abs{b}\varepsilon_s$. For $\abs{b}\varepsilon_s\geq0$, the ground state is LH and we approach the LH $g$ factor $2\kappa$ for infinite strain energy (red-dashed line). For $\abs{b}\varepsilon_s<0$, the ground state becomes more and more HH-like and we obtain the HH $g$ factor $6\kappa$ in the limit of infinite negative strain energy (blue-dashed line). 
		(b) In comparison to the results obtained without strain [cf. Fig.~\ref{fig:Heff_SOImeffGe_R15}(b)], $\alpha_{so}$ [cf. Eq.~\eqref{eqn:effSOI}] is strongly reduced. 
		The SOI is underestimated by the perturbation theory at strong electric field.
		(c) With strain, $1/\bar{m}$ [cf. Eq.~\eqref{eqn:effMassAv}] is positive also at weak $E_x$. The ground state is now more LH-like, which manifests in the fact that $1/\bar{m}$ is closer to the LH mass  $(\gamma_1+2\gamma_s)/m$ than to the average HH-LH mass  $\gamma_1/m$. 
		(d) The spin-dependence of the effective mass [cf. Eq.~\eqref{eqn:effMassDiff}] becomes less relevant with strain even at strong $B$. Note that with strain one needs to use the definitions for $c_{\pm}^s$  and $\alpha_{\pm}^s$ given by Eqs.~\eqref{eqn:cUs}, \eqref{eqn:cDs}, \eqref{eqn:alphaUs}, and~\eqref{eqn:alphaDs} in Appendix~\ref{sec:AnalyticsDerivationNW} to calculate the effective parameters. The numerics and perturbation theory disagree at $E_x =0$ due to the same reason given in the caption of Fig.~\ref{fig:Heff_SOImeffGe_R15}. Numerics and perturbation theory show both that $1/\bar{m}$ is almost constant with the electric field for the chosen value of strain due to the enhanced subband gap. \label{fig:Heff_geffSOImeffGe_str01_R15}
	}
\end{figure*}

In Ge/Si core/shell NWs, strain is a crucial feature required to increase the subband energy gap between the lowest Kramers pair  and the excited states~\cite{Kloeffel2018,Kloeffel2014,Adelsberger2022}. A large subband gap is required to define QDs because it ensures that the effective theory of $H^{2\times 2}$ defined in Eq.~\eqref{eqn:Heff2x2} is accurate even in short quantum dots with a large confinement potential along the $z$ direction. 

We describe the strain in Ge/Si core/shell NWs by the BP Hamiltonian $	H_\textrm{BP} $ [see  Eq.~\eqref{eqn:BPHam}], which is $\propto J_z^2$. As a result, we can straightforwardly extend  the analytical solution in Sec.~\ref{sec:NWAnalyticalSolution} by including the effects of strain. Strain enters the solution by modifying the coefficients $c_{\pm}^s$  and $\alpha_{\pm}^s$ in Eqs.~\eqref{eqn:cU}, \eqref{eqn:cD}, \eqref{eqn:alphaU}, and~\eqref{eqn:alphaD} to Eqs.~\eqref{eqn:cUs}, \eqref{eqn:cDs}, \eqref{eqn:alphaUs}, and~\eqref{eqn:alphaDs} given in Appendix~\ref{sec:AnalyticsDerivationNW}.

In the inset in Fig.~\ref{fig:Heff_geffSOImeffGe_str01_R15}(a), we show  how the effective $g$ factor, $g_\mathrm{eff}$, changes for different values of strain. The strain energies in the range between $\abs{b} \varepsilon_s = \SI{3.7}{\milli\electronvolt}$ and $\abs{b} \varepsilon_s = \SI{40}{\milli\electronvolt}$ can be realized by varying the relative shell thickness of the Si shell around the Ge core from $\gamma = 0.002$ to $\gamma= 0.4$~\cite{Kloeffel2014}. 
While negative values of $\abs{b} \varepsilon_s$ are not reached in Ge/Si NWs, for completeness we include these cases in our analysis. Such negative strain energies can occur in Ge NWs where the outer shell comprises a material with a larger lattice constant than Ge. In this figure, we define the $g$ factor at $B=\SI{0.01}{\tesla}$. We also remark that in the inset in Fig.~\ref{fig:Heff_geffSOImeffGe_str01_R15}(a) we plot the absolute value of  $g_\text{eff}$. In fact, interestingly, strain can cause a change of sign of $g_\text{eff}$: with positive strain energy the ground state is a $\uparrow$-state and the first excited state is a $\downarrow$-state, while with negative strain energy the order is reversed. We note that a finite value of strain (positive or negative) tends to increase  $\abs{g_\mathrm{eff}}$. This enhancement of $\abs{g_\mathrm{eff}}$ is caused by a reduced susceptibility of the NW to orbital effects. In fact, in Fig.~\ref{fig:gFacRen}, we relate the reduction of $g$ to the  avoided crossing between lowest and first Kramers pairs (blue and orange lines) induced by orbital effects. In the presence of strain, at $B=0$ the subband gap between these states is increased, thus pushing the avoided crossing to larger values of $B$, and enhancing the effective $g$ factor.

The enhanced $g$ factor at small values of $B$ is a noteworthy advantage to host MBSs and for spin qubit applications.  However, we point out that a prerequisite for the formation of MBSs is proximity-induced superconductivity in the Ge NW, a requirement that strongly limits the possible  thickness of the Si shell (and thus the values of strain). In experiments proximity-induced superconductivity was demonstrated at Si shell thicknesses between \SI{1.5}{\nano\meter} and \SI{3}{\nano\meter}~\cite{Xiang2006,Sistani2019}, corresponding to strain energies between $\abs{b} \varepsilon_s = \SI{15.7}{\milli\electronvolt}$ and $\abs{b} \varepsilon_s = \SI{26.5}{\milli\electronvolt}$ in a Ge NW with $R= \SI{15}{\nano\meter}$. As shown in Fig.~\ref{fig:Heff_geffSOImeffGe_str01_R15}(a) these strain parameters are sufficient to significantly increase the $g$ factor at weak electric field compared to the result without strain in Fig.~\ref{fig:Heff_SOImeffGe_R15}(a).

At positive strain energies, the ground state is given by $\varphi_2^\uparrow$ with the Kramers partner $\varphi_0^\downarrow$, as in the case without strain. These states are almost exclusively LH at $E_x=0$, and therefore, in the limit of infinite positive strain energy, $g_\mathrm{eff} \rightarrow 2\kappa$ [cf. red dashed line in the inset in Fig.~\ref{fig:Heff_geffSOImeffGe_str01_R15}(a)], corresponding to the pure LH $g$ factor. At negative strain, the ground state becomes $\varphi_0^\uparrow$ with the Kramers partner $\varphi_2^\downarrow$ which becomes HH-like with increasing negative strain energy. At $\abs{b} \varepsilon_s \rightarrow -\infty$ the ground state is purely HH and we obtain the HH $g$ factor $g_\mathrm{eff} \rightarrow 6 \kappa$  [cf. blue dashed line in the inset in Fig.~\ref{fig:Heff_geffSOImeffGe_str01_R15}(a)]. This trend is analogous to planar Ge heterostructures, see Sec.~\ref{sec:1dChannel}. 

We now focus on a Ge/Si core/shell NW with $\abs{b} \varepsilon_s = \SI{15.7}{\milli\electronvolt}$ ($\gamma=0.1$). We follow the calculations presented in Appendix~\ref{sec:effMod} and we adapt the formulas from Sec.~\ref{sec:2x2wireHam} to accommodate for strain, by considering the energies given by Eqs.~\eqref{eqn:dispRelGeU} and ~\eqref{eqn:dispRelGeD} with the coefficients in Eqs.~\eqref{eqn:cUs}, \eqref{eqn:cDs}, \eqref{eqn:alphaUs}, and~\eqref{eqn:alphaDs}. The results are presented in Fig.~\ref{fig:Heff_geffSOImeffGe_str01_R15}.  In comparison to the results obtained before without strain (cf. Fig.~\ref{fig:Heff_SOImeffGe_R15}), the $g$ factor is strongly enhanced at small values of $E_x$ as expected from the inset in Fig.~\ref{fig:Heff_geffSOImeffGe_str01_R15}(a).
Similar to the qualitative behavior of the $g$ factor of unstrained NWs, the perturbatively calculated $g_\mathrm{eff}$ in the presence of strain deviates form the result from numerical calculations at strong electric field, compare Figs.~\ref{fig:Heff_geffSOImeffGe_str01_R15}(a) and~\ref{fig:Heff_SOImeffGe_R15}(a).
The SOI is strongly reduced by strain [cf. Fig.~\ref{fig:Heff_geffSOImeffGe_str01_R15}(b)]. The slope in the linear regime as well as the maximum value at stronger $E_x$ is smaller than without strain because of the enhanced subband energy gap. The perturbation theory underestimates the SOI strength at strong electric field.
Moreover, the strain regularizes the inverse average effective mass $1/\bar{m}$, which is approximately constant as a function of $E_x$ and remains positive even at $E_x\rightarrow 0$, as shown in Fig.~\ref{fig:Heff_geffSOImeffGe_str01_R15}(c). The inverse mass  $1/\bar{m}$ is enlarged by strain and it approaches the LH mass $(\gamma_1+2\gamma_s)/m$  [$(\gamma_1+2\gamma_s)/\gamma_1 = 1.74$]. As can be seen from Fig.~\ref{fig:Heff_geffSOImeffGe_str01_R15}(d), strain also reduces the spin-dependent mass term. 

\section{One-Dimensional Channel} \label{sec:1dChannel}

\begin{figure*}[htb]
	\includegraphics{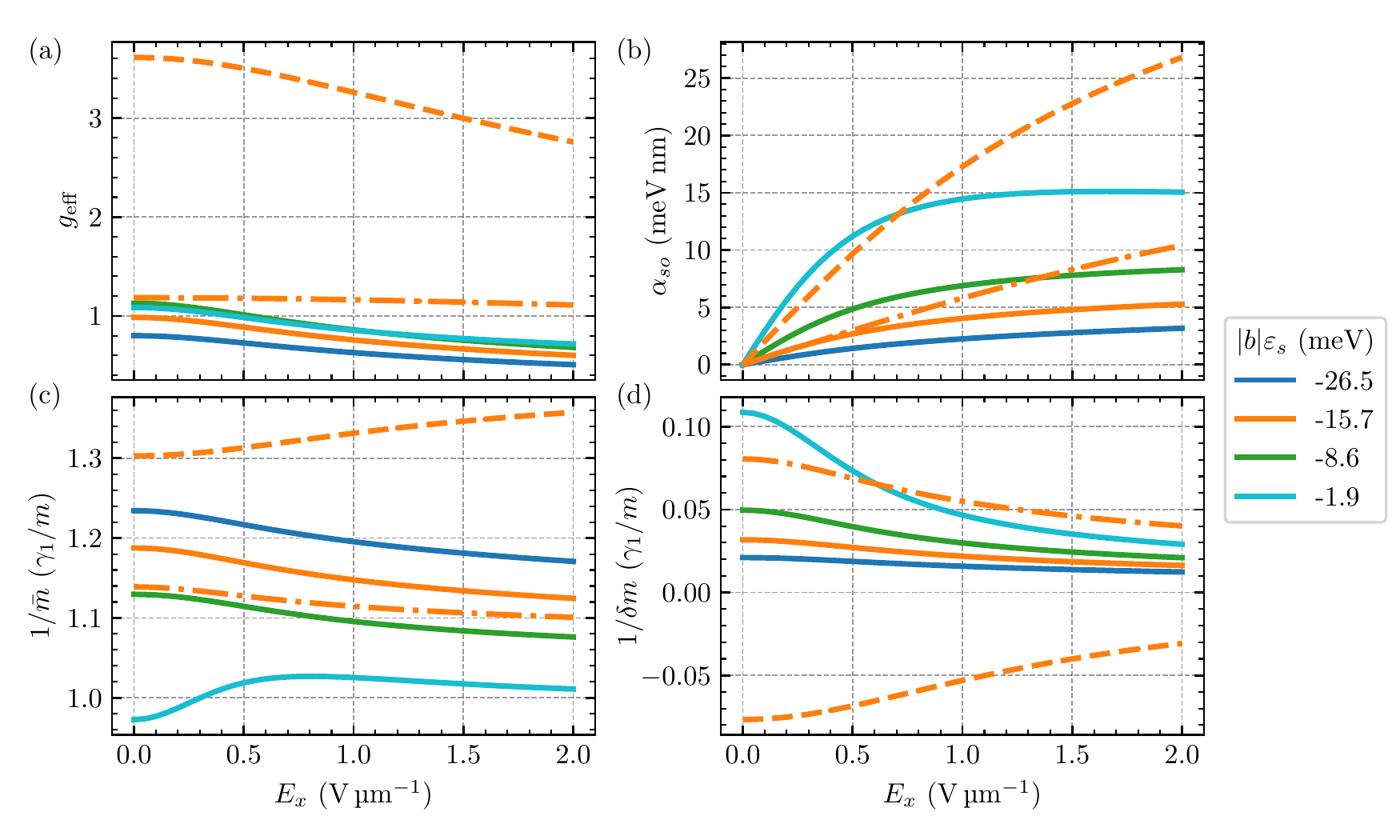}
	\caption{Effective parameters of a gate-defined one-dimensional channel in a Ge/SiGe heterostructure as a function of the electric field perpendicular to the heterostructure plane ($E_x$) with a magnetic field along the channel ($B=\SI{2}{\tesla}$). The solid lines show the numerical results (calculated as described in the main text) for the channel with a HW confinement of length $L_x = \SI{26.6}{nm}$ in $x$ direction and a harmonic confinement in $y$ direction with harmonic confinement length $l_y = L_x/\pi = \SI{8.5}{\nano\meter}$ for different values of the strain energy $\abs{b} \varepsilon_s$ indicated by the legend. The dash-dotted lines show the same for HW confinement length $L_x = \SI{15}{nm}$ for $\abs{b} \varepsilon_s = \SI{-15.7}{\milli\electronvolt}$. The dashed lines show the results for a Ge/Si core/shell NW of Radius $R=\SI{15}{\nano\meter}$ with $\abs{b} \varepsilon_s = \SI{15.7}{\milli\electronvolt}$ calculated numerically by diagonalizing the Hamiltonian in Eq.~\eqref{eqn:FullHamiltonian} discretized in real space (lattice spacing $\SI{0.25}{\nano\meter}$). Strain is included for the NW via the BP Hamiltonian  $	H_\mathrm{BP}$ from  Eq.~\eqref{eqn:BPHam} and for the channel via the BP Hamiltonian  $	H_\mathrm{BP}^\mathrm{ch}$ from Eq.~\eqref{eqn:BPHam_channel}. The $g$ factor and SOI of the channel exhibit the same qualitative behavior as the NW, while the mass terms are different in the two different geometries.
		(a) The effective $g$ factor of the channel is much smaller than for the NW. A decrease of the channel cross section (dash-dotted line) leads to a weaker dependence on the external electric field and a slightly larger $g$ factor.
		(b) The SOI of the channel with $L_x= \SI{26.6}{nm}$ and $\abs{b} \varepsilon_s = \SI{-1.9}{\milli\electronvolt}$ is quantitatively in the same range as the SOI of the NW at weak electric field. With increasing strain in the channel, the SOI decreases.
		(c) While the NW exhibits an inverse average effective mass that increases, the same quantity decreases with $E_x$ in the channel except for  $\abs{b} \varepsilon_s = \SI{-1.9}{\milli\electronvolt}$ where a maximum occurs at $E_x = \SI{0.8}{\volt\per\micro\meter}$. Generally, $1/\bar{m}$ is smaller in the channel than in the NW.
		(d) The inverse spin-dependent mass has an opposite sign in the channel with respect to the NW.  \label{fig:Channel_Heff_SOImeffGe_L15comp}}
\end{figure*}

In  this section, we analyze how the details of the confinement affect the parameters of the effective model, $H^{2\times 2}$, introduced in Eq.~\eqref{eqn:Heff2x2}. In particular we consider a one-dimensional channel defined by gates in a planar Ge/SiGe heterostructure as shown in Fig.~\ref{fig:Three1dSystems}(b). The channel extends in the $z$ direction. We consider a HW confinement, see Eq.~\eqref{eqn:HW_boundaryCondition},  in $x$ direction, perpendicular to the substrate, that models the interfaces between the Ge layer with width $L_x$ and the SiGe layers. We consider an electrostatic gate creating a harmonic confinement potential in the $y$ direction given by
\begin{align}
	U(y) = \frac{\hbar \gamma_1}{2 m l_y^4}y^2.
\end{align}
The confinement potential is parametrized by the harmonic confinement length $l_y$. Here,  we consider $l_y= L_x/\pi$ in order to have comparable confinement in $x$ and $y$ direction. The magnetic field is applied in the $z$ direction, parallel to the channel.
Strain is included via the BP Hamiltonian $	H_\mathrm{BP}^\mathrm{ch}$  given in Eq.~\eqref{eqn:BPHam_channel}. For our calculations we choose realistic values for the strain energy compared to values typically measured in Ge/SiGe heterostructures ($|b|\varepsilon_s\approx \SI{16}{\milli\electronvolt}$)~\cite{Sammak2019} and we also analyze the limit of weak strain. 

In this architecture, we solve the problem by diagonalizing the Hamiltonian in Eq.~\eqref{eqn:FullHamiltonian} at $k_z=0$ numerically directly including the electric field.
For the numerical diagonalization of the LK Hamiltonian we use the first 20 eigenstates of the harmonic oscillator in $y$ direction and the basis 
\begin{align}
	f_{n_x} (x) = \frac{\sqrt{2}\sin\left[n_x \left(\frac{x}{L_x}+\frac{1}{2}\right)\right]}{\sqrt{L_x}}
\end{align}
with $0< n_x \leq 20$ in $x$ direction, fulfilling the HW boundary conditions.

In Fig.~\ref{fig:Channel_Heff_SOImeffGe_L15comp}, we present the results of our analysis where we consider a magnetic field  of $B=\SI{2}{\tesla}$ along the channel and a homogeneous electric field $E_x$ perpendicular to the substrate. We compare the results for a channel of HW confinement length $L_x = \SI{26.6}{nm}$ and harmonic confinement length $l_y = L_x/\pi = \SI{8.5}{\nano\meter}$ (solid lines) for different values of the strain energy to a Ge/Si core/shell NW of radius $R=\SI{15}{\nano\meter}$ and strain energy $\abs{b} \varepsilon_s = \SI{15.7}{\milli\electronvolt}$ (dashed lines). With these choices for the confinement details, the areas of the cross sections in the two cases are comparable. We emphasize again that the sign of the strain energy in planar Ge is opposite to the strain in the NW, see Eqs.~\eqref{eqn:BPHam} and~\eqref{eqn:BPHam_channel}. 
Furthermore, we also make a comparison to a channel with a much smaller cross section with $L_x = \SI{15}{nm}$ and $l_y = \SI{15}{nm}/\pi = \SI{4.8}{\nano\meter}$ (dot-dashed lines).

The one-dimensional channel geometry exhibits a few features that are different from those observed in the NW. In particular, as shown in Fig.~\ref{fig:Channel_Heff_SOImeffGe_L15comp}, only the SOI and the $g$ factor exhibit the same qualitative behavior in both the NW and channel geometry. In a channel with $\abs{b} \varepsilon_s = \SI{-1.9}{\milli\electronvolt}$, we also observe at weak electric field quantitatively similar values of $\alpha_{so}$ as in a NW with $\abs{b} \varepsilon_s = \SI{15.7}{\milli\electronvolt}$. In Fig.~\ref{fig:Channel_Heff_SOImeffGe_L15comp}(a), we show that $g_\mathrm{eff}$ is significantly smaller in the channel  than in the NW and that it decreases with the amount of strain in the planar structure. In the channel with smaller cross section, the electric field dependence of $g_\mathrm{eff}$ is reduced.

In Figs.~\ref{fig:Channel_Heff_SOImeffGe_L15comp}(c) and~\ref{fig:Channel_Heff_SOImeffGe_L15comp}(d), we analyze the inverse average and spin-dependent masses, respectively. With increasing (negative) strain energy the inverse average mass $1/\bar{m}$ increases. Also the qualitative behavior changes and instead of  
having a maximum that occurs for $\abs{b} \varepsilon_s = \SI{-1.9}{\milli\electronvolt}$, at larger strain energies, $1/\bar{m}$ decreases monotonically with $E_x$. This trend is in contrast to the monotonic increase of the inverse averages mass in the Ge/Si NW. Interestingly, the inverse spin-dependent mass is positive in the channel, while it is negative for the NW with respect to the $g$ factor. As in the NW geometry, also in the channel geometry the spin-dependent mass is most relevant at $E_x \rightarrow 0$ and for small values of strain, and it is negligible otherwise.
 
\section{Curved quantum well} \label{sec:thinShell}

In the following, we consider a CQW as sketched in Fig.~\ref{fig:Three1dSystems}(c). 
 The confinement for the CQW is given by
\begin{align}
	V(r) = 
	\begin{cases}
		\infty, &r = \sqrt{x^2 + y^2}  < R,\\ 
		0, & R < r  < R_1,\\ 
		\infty, &  r > R_1
	\end{cases} \label{eqn:HW_boundaryConditionHollowWire}
\end{align}
where the radii are defined as in Fig.~\ref{fig:Three1dSystems}(c). 
The strain induced into the Ge shell is modeled by the BP Hamiltonian $	H_\mathrm{BP}^\mathrm{CQW}$ defined in Eq.~\eqref{eqn:HBP_ts}, as discussed in Ref.~\cite{Bosco2022}.
The longitudinal and radial strain energies can be engineered individually by the radii of the inner and outer shell, $R_1$ and $R_2$, see Eqs.~\eqref{eqn:strainLong} and~\eqref{eqn:strainRad}.
Only the longitudinal strain energy $\abs{b}\varepsilon_z$ depends on the thickness of the outer shell. 

\begin{figure}[htb]	\includegraphics{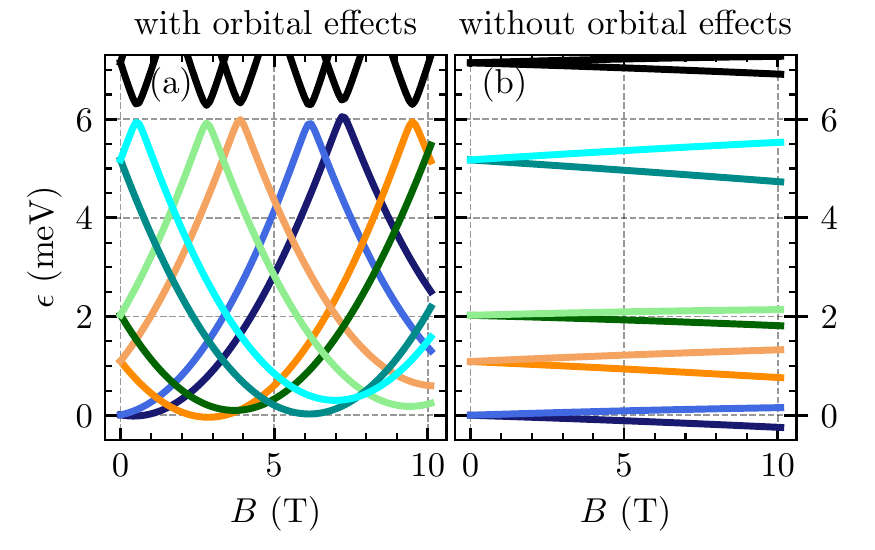}
	\caption{Energy spectrum of a CQW  as a function of the magnetic field $B$ at $k_z=0$ calculated by numerically diagonalizing  $H_\mathrm{ILK} + H_\mathrm{BP}^\mathrm{CQW} + H_Z + V$ (a) including orbital effects and (b) neglecting orbital effects. Orbital effects have a huge impact on the evolution of the energy states with the magnetic field. Only at weak $B<\SI{1}{\tesla}$ the levels in the panel (a) do not cross. The colors indicate Kramers partners. The inner core radius is  given by $R=\SI{15}{\nano\meter}$. The thin outer Ge shell is defined by $R_1=\SI{25}{\nano\meter}$ and $R_2=\SI{35}{\nano\meter}$. We use a lattice spacing of \SI{0.5}{\nano\meter}. \label{fig:Spec_E0_Rin15Rout25Rs35_str}}
\end{figure}

\begin{figure*}[htb]	\includegraphics{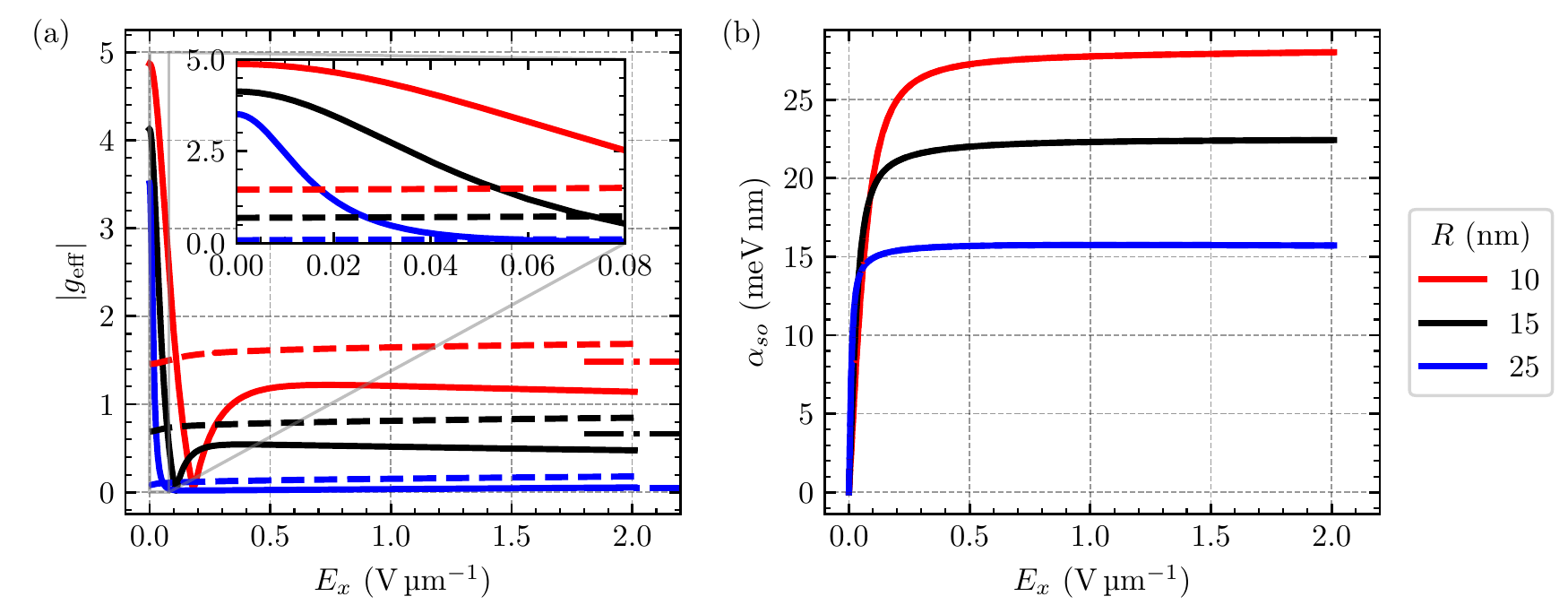}
	\caption{(a)  Effective $g$ factor $g_\mathrm{eff}$ and (b) SOI $\alpha_{so}$  of a CQW ($R_1-R=\SI{10}{\nano\meter}$ and $R_2-R_1=\SI{10}{\nano\meter}$) as a function of the electric field $E_x$ at $k_z=0$ and $B=\SI{0.1}{\tesla}$ calculated by numerically diagonalizing the Hamiltonian in Eq.~\eqref{eqn:FullHamiltonian} accounting for strain via $H_\mathrm{BP}^\mathrm{CQW}$ defined  in Eq.~\eqref{eqn:HBP_ts} and the homogeneous electric field via $H_E = -e E_x x$. (a) The solid (dashed) lines correspond to the results calculated including (excluding) orbital effects.  At weak $E_x \lesssim \SI{0.05}{\volt\per\micro\meter}$, the lowest energy states are close and influence each other strongly. If  $E_x$ is increased, the subband gap between the ground state Kramers pair and the excited states also increases, resulting in a constant $g$ factor. At moderate electric field, $g_\mathrm{eff}$ is smaller when orbital effects are taken into account. The dash-dotted lines in the right of the panel (a) correspond to $g_\mathrm{eff}$ obtained from Eq.~\eqref{eqn:CQW_AnaGeff}, which is valid at strong $E_x$. (b) We find that $\alpha_{so}$ is the same with and without orbital effects. The SOI increases rapidly at weak $E_x$ and then stays constant at a large value.  We use a lattice spacing of \SI{0.5}{\nano\meter}. \label{fig:EffPar_B01_Rs10_str_orbNoOrb}}
\end{figure*}

We now calculate the energy spectrum of the CQW as a function of the magnetic field by numerically diagonalizing the discretized version of the Hamiltonian in Eq.~\eqref{eqn:FullHamiltonian} where we account for the BP Hamiltonian $H_\mathrm{BP}^\mathrm{CQW}$ defined  in Eq.~\eqref{eqn:HBP_ts} and the confinement introduced above.
Orbital effects are crucial in the CQW as illustrated by Fig.~\ref{fig:Spec_E0_Rin15Rout25Rs35_str} and discussed in Ref.~\cite{Bosco2022}. A comparison between Fig.~\ref{fig:Spec_E0_Rin15Rout25Rs35_str}(a) [with orbital effects] and Fig.~\ref{fig:Spec_E0_Rin15Rout25Rs35_str}(b) [without orbital effects] clearly illustrates the strikingly different magnetic field dependence of the system properties when orbital effects are included. 
In particular, there is a number of level crossings when orbital magnetic field is accounted for. Only at weak magnetic field ($B<\SI{1}{\tesla}$), the levels in Fig.~\ref{fig:Spec_E0_Rin15Rout25Rs35_str}(a) do not cross, thus, in the following we focus on the regime of weak magnetic field. 

The states in the CQW are close in energy even with strain~\cite{Bosco2022}, thus, we cannot accurately include an electric field perturbatively. As a result, we include a homogeneous electric field perpendicular to the NW numerically. The results we obtain by this approach are presented in Fig.~\ref{fig:EffPar_B01_Rs10_str_orbNoOrb}. We compare the effective $g$ factor and the SOI strength for different radii $R$ as well as  with and without including orbital effects. We consider a Ge shell of width $R_1-R=\SI{10}{\nano\meter}$ and the outer Si shell $R_2-R_1=\SI{10}{\nano\meter}$. The strain energies for the different radii considered in Fig.~\ref{fig:EffPar_B01_Rs10_str_orbNoOrb} are provided in Table~\ref{tab:strainEnergies}. With increasing core radius $R$ the longitudinal strain component $\varepsilon_z$ decreases quickly and becomes negligible at $R= \SI{25}{\nano\meter}$. In contrast, the radial strain component $\varepsilon_r$ increases with the core radius.

\begin{table}[!b]
	\caption{Longitudinal ($\abs{b}\varepsilon_z$) and radial  ($\abs{b}\varepsilon_r$) strain energy in the CQW according to Eq.~\eqref{eqn:strainLong} and Eq.~\eqref{eqn:strainRad}, respectively, for the inner radii chosen for Fig.~\ref{fig:EffPar_B01_Rs10_str_orbNoOrb}. The Ge shell thickness is fixed at $R_1-R=\SI{10}{\nano\meter}$ and the outer shell radius is fixed at $R_2-R_1=\SI{10}{\nano\meter}$. \label{tab:strainEnergies}}
	\begin{ruledtabular}
		\begin{tabular}{c|ccc}
			$R$ (\si{\nano\meter})  & $10$ &  $15$ & $25$\\
			$\abs{b}\varepsilon_z$ (meV) & $15.6$ & $7.8$ & $0.65$\\
			$\abs{b}\varepsilon_r$ (meV)  &  $62.6$ & $79.2$ & $97.8$
		\end{tabular}
	\end{ruledtabular}
\end{table}

Importantly, a finite electric field causes an increase of the small subband gap between the ground state Kramers pair and the excited states shown in Fig.~\ref{fig:Spec_E0_Rin15Rout25Rs35_str}~\cite{Bosco2022}. Due to the small subband gap at weak electric field, the effective $g$ factor in Fig.~\ref{fig:EffPar_B01_Rs10_str_orbNoOrb}(a) changes rapidly.  It becomes a rather  constant function of the electric fields above a certain critical field. An almost electric-field-independent $g_\mathrm{eff}$  is a critical advantage for spin qubit applications because it strongly suppresses the susceptibility to charge noise~\cite{Vion2002,Petersson2010}, a key issue in hole NWs~\cite{Froning2021a,Wang2022}. 
We anticipate that the large $g$ factor for $R= \SI{10}{\nano\meter}$ can acquire significant corrections coming from the  high energy holes, as we will discuss in Sec.~\ref{sec:corrections}. Despite the weak magnetic field of $B=\SI{0.1}{\tesla}$, the $g$ factor is enhanced considerably at weak electric field ($E_x < \SI{0.1}{\volt\per\micro\meter}$) due to orbital effects. For small radius, the orbital effects reduce the $g$ factor at strong electric field.
The main reason for the $g$ factor to decrease with increasing radius $R$ is not the weaker confinement in the larger cross section but the larger value of radial strain (cf. Table~\ref{tab:strainEnergies}).

Analytical calculations analogous to the ones in Ref.~\cite{Bosco2022} predict that the $g$ factor at strong electric fields is independent of  orbital effects and reduces to 
\begin{align}
	g_\mathrm{eff} = 6 \kappa \frac{\abs{b}\varepsilon_z}{\abs{b}\varepsilon_z + 2 \abs{b}\varepsilon_r + \frac{2\hbar^2\pi^2\gamma_s}{m (R_1-R)^2}}. \label{eqn:CQW_AnaGeff}
\end{align}
We show the predicted value as dot-dashed lines on the right side of Fig.~\ref{fig:EffPar_B01_Rs10_str_orbNoOrb}. The analytical formula provides a good estimation for the $g$ factor for thin Ge shells and large values of $\abs{b}\varepsilon_z$ but we observe that it also reasonably captures the $g$ factor at rather small values of longitudinal strain and thick shells. In the latter cases, there are small variations of $g_\mathrm{eff}$ by orbital effects. These variations however do not change the slope of the curves and $g_\mathrm{eff}$ remains a rather flat function of $E_x$. 

In contrast to the $g$ factor, the SOI strength [cf. Fig.~\ref{fig:EffPar_B01_Rs10_str_orbNoOrb}(b)] is not influenced by the orbital magnetic field at this value of the magnetic field for any of the chosen values of $R$. At weak electric field the SOI increases rapidly and then remains constant at a large value. This behavior is extremely advantageous for various spin qubit  applications because it removes the need of fine-tuning the electric field to reach the maximal value of $\alpha_{so}$.  The SOI could still be switched off  completely at $E_x=0$.  In contrast to the $g$ factor, the SOI depends mainly on the radius and less on the strain energy. We note that even at weak electric field where the $g$ factor remains sizable the SOI can also be large (at $E_x = \SI{0.06}{\volt\per\micro\meter}$ and $R=\SI{10}{\nano\meter}$: $g_\mathrm{eff} = 3.3$ and $\alpha_{so}=\SI{15}{\milli\electronvolt\nano\meter}$), which makes this platform suitable to find Majorana bound states.

\section{Corrections to the Model} \label{sec:corrections}

In this section, we resort to fully numerical calculations using a discretized model in real space  to analyze the validity of the analytical results in the presence of additional effects including split-off holes and cubic anisotropies. We focus here on cylindrical Ge/Si core/shell NWs and on CQWs.

\subsection{Spin-orbit split-off band}
In the following we explore the effect of  the spin-orbit SOB on the effective $g$ factor, the SOI, and the effective mass of Ge NWs and CQWs. We calculate the effective parameters with a $6\times6$  LK Hamiltonian (cf. Appendix~\ref{sec:6bandLKHam}) taking into account the two LH, the two HH and the two split-off hole states. To study the effect of the SOB, here, we restrict our analysis to the isotropic $6\times6$  LK Hamiltonian ($\gamma_2=\gamma_3=\gamma_s$). The spin-orbit gap for Ge is $\Delta_{SO} =\SI{296}{\milli\electronvolt}$~\cite{Winkler2003}. However, despite this large gap, the SOB renormalizes the parameters of the effective model and causes a considerable quantitative change in the system. Note that the spin of the split-off holes is truly $1/2$ while the LHs correspond to the $\pm 1/2$ eigenvalues of the spin-$3/2$ matrix $J_z$.

\subsubsection{Ge nanowire}

\begin{figure}[htb]
	\includegraphics{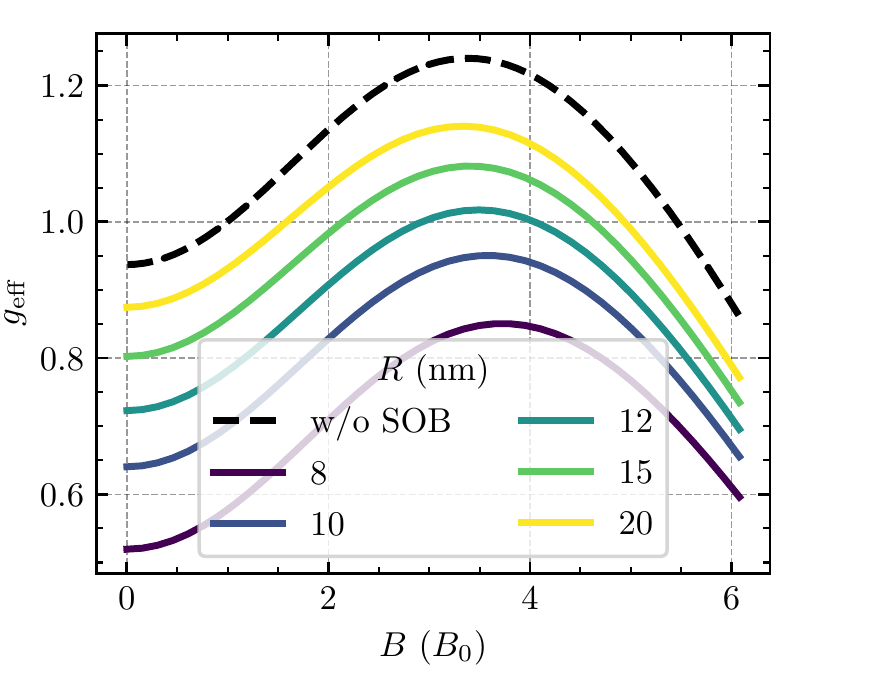}
	\caption{Effective $g$ factor $g_\mathrm{eff}$ of a Ge NW with circular cross section as a function of the magnetic field (in units of $B_0 = \SI{658.2}{\tesla\times\nano\meter\squared\per R \squared}$) calculated numerically using the $6\times6$  LK Hamiltonian $H^{6\times 6}$ (given in Appendix~\ref{sec:6bandLKHam}), which takes into account the spin-orbit SOB. For comparison we provide $g_\mathrm{eff}$ without the spin-orbit SOB (dashed line) calculated semi-analytically according to  Eqs.~\eqref{eqn:dispRelGeU}, ~\eqref{eqn:dispRelGeD}, and~\eqref{eqn:effgFac}. The spin-orbit SOB causes a decrease of $g_\mathrm{eff}$ that  is larger at small $R$. We use a lattice spacing of \SI{0.5}{\nano\meter}.\label{fig:geff_GeAna_SOband}}
\end{figure}

\begin{figure*}[htb]
	\includegraphics{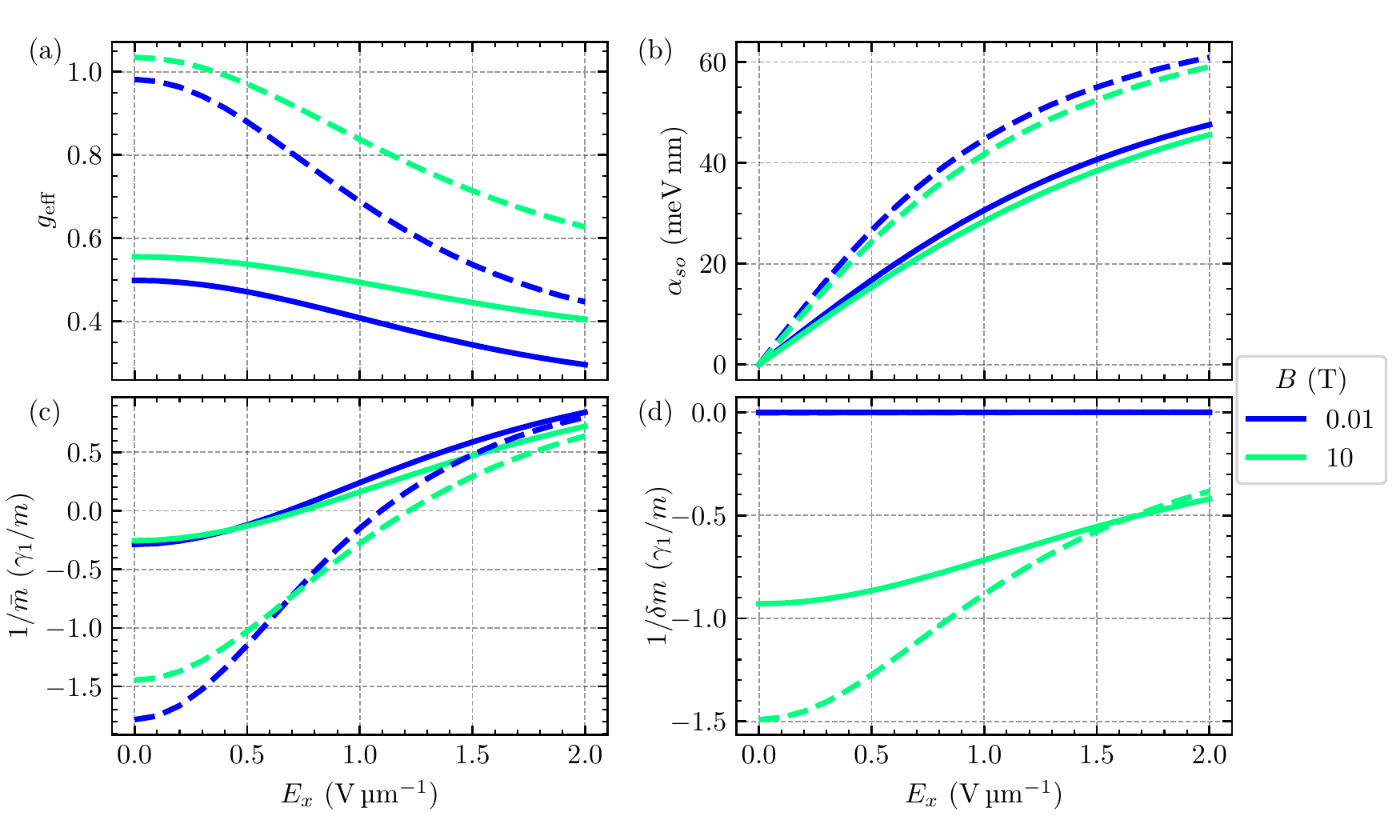}
	\caption{Effective parameters of a Ge NW with circular cross section as a function of the electric field $E_x$ calculated numerically with the $6\times6$ LK Hamiltonian $H^{6\times 6}$ (given in Appendix~\ref{sec:6bandLKHam}), which takes into account the spin-orbit SOB (solid lines). For comparison we provide the result without the spin-orbit SOB (dashed lines). We choose a relatively small radius $R=\SI{8}{\nano\meter}$ since the influence of the SOB is stronger for stronger confinement. (a) As we already know from Fig.~\ref{fig:geff_GeAna_SOband}, the effective $g$ factor is reduced when the SOB is included. However, the qualitative $E_x$ dependence stays the same. (b) The spin-orbit SOB reduces $\alpha_{so}$ only slightly. The correction is smallest at weak electric field. (c) At weak $E_x$, $\abs{1/\bar{m}}$ is reduced due to the influence of the SOB. In addition, with the SOB, the average mass is negative at weak and positive at strong electric fields. (d) The SOB influences  $\abs{1/\delta m}$ only slightly towards a smaller value. As for the average mass the correction is largest at weak $E_x$. For all effective parameters, the difference between $B=\SI{0.1}{\tesla}$ and  $B=\SI{10}{\tesla}$ is small in agreement with the results presented in Fig.~\ref{fig:Heff_geffSOImeffGe_str01_R15}. We use a lattice spacing of \SI{0.16}{\nano\meter}. \label{fig:Heff_SOImeffGe_R8_SOB}}
\end{figure*}

Our result from numerical calculations of the $g$ factor of a Ge NW is presented in Fig.~\ref{fig:geff_GeAna_SOband}. 
Due to the SOB, the effective $g$ factor at $E_x=0$ depends on both  $R/l_B$ and $\Delta_{SO}/\hbar\omega_c$. 
By comparing with Fig.~\ref{fig:gFacRen}(c), we observe that qualitatively the dependence of $g_\text{eff}$ on $B$ still resembles the one obtained for the $4\times 4$ LK Hamiltonian but  the SOB tends to reduce $g_\text{eff}$, especially in NWs with a small radius. With increasing radius the $4\times 4$ LK Hamiltonian becomes more accurate because  the confinement energy $\propto 1/R^2$ becomes smaller compared to $\Delta_{SO}$, and the states in the SOB are well-separated from the low-energy HH-LH subspace. 
More precisely, in Table~\ref{tab:maxGeff_GeB6x6}, we show the dependence of the position (at $B=\tilde{B}$) and value 
[$g_\text{eff}(B=\tilde{B})$] of the maximum of $g_\mathrm{eff}$ for the five radii used in Fig.~\ref{fig:geff_GeAna_SOband}.
 At $R= \SI{20}{\nano\meter}$ the maximum value of $g_\mathrm{eff}$ deviates only $8\%$ from the value of  $1.24$ obtained without accounting for the SOB, thus justifying the analysis in Sec.~\ref{sec:AnaSolNW}.

\begin{table}[!b]
	\caption{The position ($\tilde{B}$) and value ($g_\mathrm{eff, max}$) of the maximum of the effective $g$ factor in Ge NWs with circular cross section for different radii obtained numerically to include the SOB. \label{tab:maxGeff_GeB6x6}}
	\begin{ruledtabular}
		\begin{tabular}{c|ccccc}
			Radius $R$ (\si{\nano\meter})  & $8$ & $10$ & $12$ & $15$ & $20$\\
			$\tilde{B}$ (T) & $39.1$ & $24.0$ & $16.0$ & $9.8$ & $5.5$ \\
			$g_\mathrm{eff, max}$  &  $0.85$ & $0.95$ & $1.02$ & $1.08$ & $1.14$
		\end{tabular}
	\end{ruledtabular}
\end{table}

We study the effect of an electric field by numerically diagonalizing the discretized version of the Hamiltonian in Eq.~\eqref{eqn:FullHamiltonian} and show the results in Fig.~\ref{fig:Heff_SOImeffGe_R8_SOB}.
We compare numerical results obtained numerically by including the SOB with the results obtained from the $4\times 4$ ILK Hamiltonian $H_\mathrm{ILK}$ [cf. Eq.~\eqref{eqn:LK_Hamiltonian_sphericalApp}]. The holes in the SOB are more effective at small values of $R$. Thus, to emphasize their effect, here, we present the results of a simulation of a NW of a small radius $R=\SI{8}{\nano\meter}$.  
The SOB does not alter the qualitative behavior of the effective parameters, however, these states can renormalize the values quantitatively.
In particular, in Fig.~\ref{fig:geff_GeAna_SOband}(a) and (b), we show that $g_\text{eff}$ is reduced by the SOB also at finite $E_x$ and that the effective SOI $\alpha_{so}$ is only weakly renormalized even in narrow NWs, respectively. 
The effective masses also acquire corrections because of the SOB, which are  more pronounced at small values of $E_x$, see Figs.~\ref{fig:Heff_SOImeffGe_R8_SOB}(c) and (d).
At stronger electric fields, the influence of the SOB on both mass terms becomes negligible. 

Our findings confirm that, in most cases, the SOB only causes a quantitative correction to the $4\times 4$ LK Hamiltonian defined for a Ge NW. This correction is rather small in wide NWs but it can be significant in narrow NWs and should be included in these cases to have an accurate description of the system.

\begin{figure*}[htb]	\includegraphics{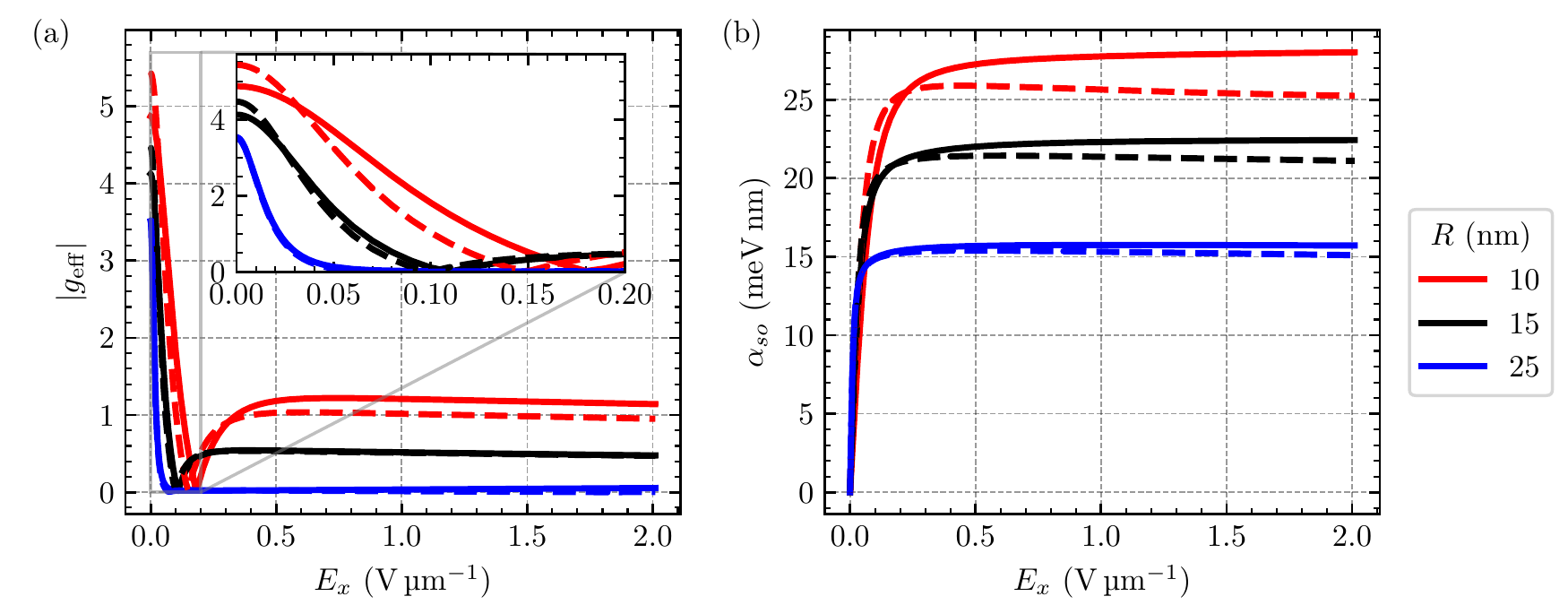}
	\caption{Numerically calculated  (a)  effective $g$ factor $g_\mathrm{eff}$ and  (b) SOI strength $\alpha_{so}$ of a CQW with $R_1-R=\SI{10}{\nano\meter}$ and $R_2-R_1=\SI{10}{\nano\meter}$ [cf. Fig.~\ref{fig:Three1dSystems}(c)] as a function of the electric field $E_x$ at $k_z=0$ and $B=\SI{0.1}{\tesla}$. The dashed lines depict the results calculated with the $6\times 6$ LK Hamiltonian $H^{6\times 6}$ given in Appendix~\ref{sec:6bandLKHam}, which includes the spin-orbit SOB.  The solid lines correspond to the results obtained without including the SOB. Only for strong confinement ($R=\SI{10}{\nano\meter}$) the split-off holes renormalize both  the $g$ factor and the SOI strength. For weaker confinement the influence of the split-off holes is negligible. We use a lattice spacing of \SI{0.5}{\nano\meter}. \label{fig:EffPar_B01_Rs10_str_SOB}}
\end{figure*}

\subsubsection{Curved quantum well}

The effect of the SOB on the $g$ factor and the SOI  strength of CQWs is shown by Fig.~\ref{fig:EffPar_B01_Rs10_str_SOB}. In order to account for strain, in our numerical calculations, we resort to the $6\times 6$ BP Hamiltonian given in Appendix~\ref{sec:6bandLKHam}, see Eq.~\eqref{eqn:6x6BPHam}. Again, we compare the results obtained with the $4\times 4$ Hamiltonian neglecting the SOB (solid lines) to the results where the SOB is accounted for (dashed lines). 
In analogy to Ge/Si core/shell NWs, in CQWs, the effects of the SOB are strongest in NWs with small radii, resulting in a renormalization of $g_\text{eff}$ by up to $20\%$ at $R= \SI{10}{\nano\meter}$ and $E_x = \SI{1.0}{\volt\per\micro\meter}$. This correction is comparable to what we observe in the Ge NW of radius $R= \SI{8}{\nano\meter}$ [cf. Fig.~\ref{fig:Heff_SOImeffGe_R8_SOB}(a)]. For weaker confinement and larger $R$ the renormalization due to SOB becomes negligible. The SOB affects the SOI [cf. Fig.~\ref{fig:EffPar_B01_Rs10_str_SOB}(b)] in a similar way.

\subsection{Cubic Luttinger-Kohn anisotropies}
In this subsection, we discuss the effects of anisotropy on the effective parameters of a Ge/Si core/shell NW and of a CQW. In addition,  we investigate the validity of the ILK Hamiltonian $H_\mathrm{ILK} $ defined in Eq.~\eqref{eqn:LK_Hamiltonian_sphericalApp}. 
Here, we calculate the effective parameters numerically by using the general LK Hamiltonian $H_\mathrm{LK} $  provided in Eq.~\eqref{eqn:LK_Hamiltonian}.We focus on rather wide NWs, thus, we neglect the SOB. If we include the cubic anisotropies of the LK Hamiltonian, the growth direction of the NW becomes relevant~\cite{Kloeffel2018,Bosco2021b}. Here, we consider the three situations introduced in Sec.~\ref{sec:ModelOfNW}, where the NW is grown along $z\parallel [001]$, $z\parallel [110]$, and $z\parallel [111]$, and compare these cases to the results obtained from the ILK Hamiltonian. 

\subsubsection{Ge/Si core/shell nanowire}

\begin{figure}[htb]
	\includegraphics{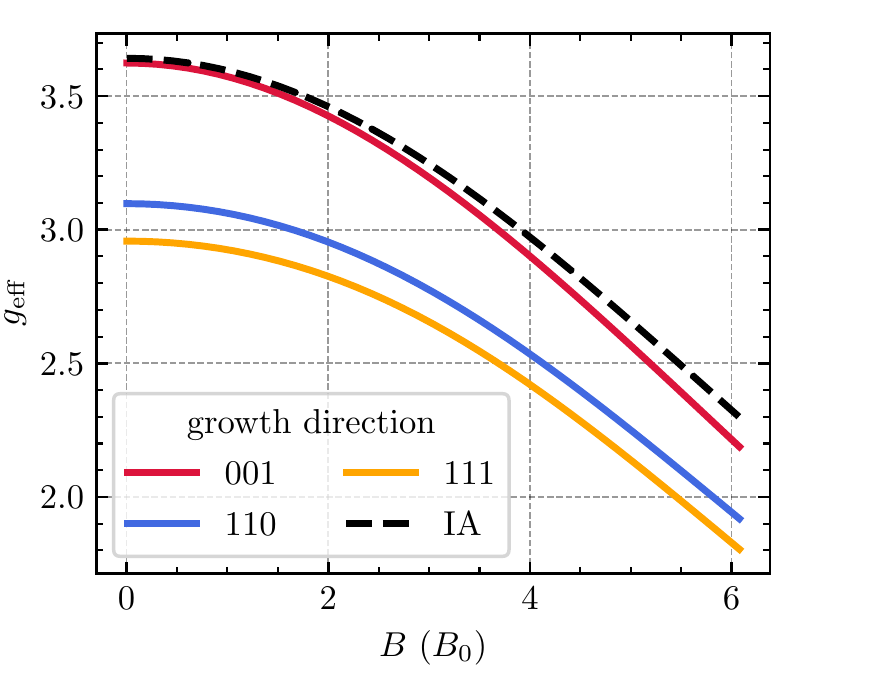}
	\caption{Effective $g$ factor $g_\mathrm{eff}$ of a strained Ge/Si core/shell NW with circular cross section as a function of $B$ (in units of $B_0 = \SI{658.2}{\tesla\times\nano\meter\squared\per R \squared}$) calculated numerically for different growth directions. Here, we fix strain to $\abs{b}\varepsilon_s = \SI{15.7}{\milli\electronvolt}$. For comparison, we also provide the result obtained from the ILK Hamiltonian (black dashed line). The $[001]$ NW growth direction matches the isotropic result well. The growth directions $[110]$ and $[111]$ exhibit the same qualitative behavior of  the $g$ factor as the isotropic result. However, quantitatively for these growth directions $g_\mathrm{eff}$ is smaller. We use a lattice spacing of \SI{0.5}{\nano\meter}. \label{fig:geff_Ge_growthDir_R15_str01}}
\end{figure}

In Fig.~\ref{fig:geff_Ge_growthDir_R15_str01}, we show  $g_\text{eff}$ at $E_x=0$  in a Ge/Si core/shell NW as a function of the magnetic field applied parallel to the NW axis. Comparing to the result obtained within the ILK approximation (black-dashed line), we observe that the anisotropies reduce the $g$ factor, especially, when the $z$ axis is not aligned to a main crystallographic axis. In fact, when $z\parallel[001]$, the $g$ factor agrees well with the ILK but, at $z\parallel[110]$ or $z\parallel[111]$, $g_\text{eff}$ is significantly smaller. For a more anisotropic material such as Si, we expect a larger difference between different growth directions but we do not analyze this case here.

\begin{figure*}[htb]
	\includegraphics{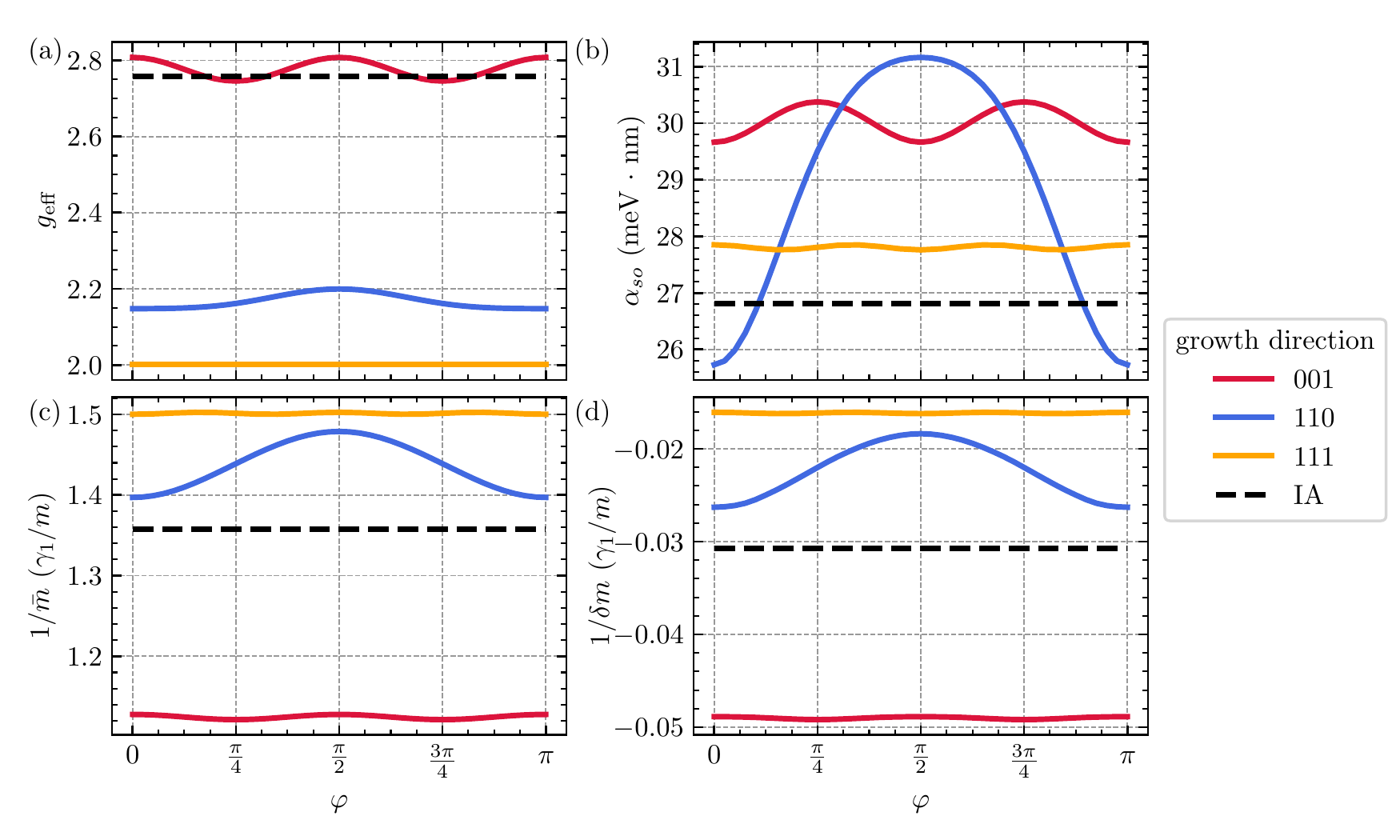}
	\caption{Effective parameters of a  strained Ge/Si core/shell NW with circular cross section of radius $R=\SI{15}{\nano\meter}$ as a function of the electric field direction represented by the angle $\varphi$ calculated numerically ($B=\SI{2}{\tesla}$, $E = \SI{2}{\volt\per\micro\meter}$, and $\abs{b}\varepsilon_s = \SI{15.7}{\milli\electronvolt}$) by diagonalizing the discretized version of the Hamiltonian in Eq.~\eqref{eqn:FullHamiltonian}. The NW growth direction is fixed  to  $z\parallel [110]$ with $\vect{E}\perp z$. For $\varphi = 0$, $\vect{E}\parallel [001]$, whereas, for $\varphi = \pi/2$, $\vect{E}\parallel [\bar{1}10]$, and, for $\varphi = \pi$, $\vect{E}\parallel [00\bar{1}]$. Except for the SOI for the $[110]$ growth direction the parameters depend only weakly on the growth direction of the NW. 
		 (a) Regardless of the growth direction of the NW the $g$ factor is to a good approximation independent of the electric field direction. Quantitatively the IA gives a good estimate for the $g$ factor of a NW grown along one of the main crystallographic axes. For the growth directions $[110]$ and $[111]$ the $g$ factor is smaller than expected from the IA.
		 (b) Only for the $[110]$ growth direction the SOI depends on the direction of the electric field significantly and we find a maximum at $\vect{E}\parallel  [\bar{1}10]$. The IA gives a good estimate for the average SOI. 
		 [(c),(d)] The effective mass terms depend only slightly on the direction of the electric field and agree on average well with the results from the isotropic LK Hamiltonian. As expected at strong electric field the spin-dependent mass term approaches zero.
		 \label{fig:effpar_R15_E2phi_comp}}
\end{figure*}

In the following, we include a homogeneous electric field perpendicular to the NW axis and diagonalize the Hamiltonian in Eq.~\eqref{eqn:FullHamiltonian} disrectized in  real space. As we account for anisotropies, the effective parameters  depend on the direction of the electric field. In Fig.~\ref{fig:effpar_R15_E2phi_comp} we compare the results for the effective parameters of NWs of the three growth directions $[001]$, $[110]$, and $[111]$ and an isotropic NW as a function of the direction of the electric field. We plot the results at $B=\SI{2}{\tesla}$ and at a rather strong electric field, $E = \SI{2}{\volt\per\micro\meter}$, where we expect a large effect of the anisotropies. However, for all three growth directions the parameters only weakly depend on the direction of the electric field, suggesting that the IA is a good approximation to describe Ge NWs even at strong electric fields. The effective $g$ factors for the $[110]$ and $[111]$ growth directions deviate quantitatively from the result obtained with the isotropic LK Hamiltonian. For the NW grown parallel to the $[110]$ direction the SOI depends on the direction of the electric field with a maximum of $\alpha_{so} = \SI{31.2}{\milli\electronvolt \nano\meter}$ at $\vect{E}\parallel  [\bar{1}10]$ ($\varphi = \pi/2$). The effective mass terms are well described by the IA. As expected from Sec.~\ref{sec:strain},  the inverse average effective mass is positive because we are analyzing a strained Ge NW. In addition, it is larger than the average HH-LH mass $\gamma_1/m$. We find that $1/\delta m$ is small and negative, in agreement with the isotropic results shown in Fig.~\ref{fig:Heff_geffSOImeffGe_str01_R15}(d).

In summary, the ILK Hamiltonian $H_\mathrm{ILK} $  is well suited to describe the behavior of Ge/Si core/shell NWs grown along the [001], [110], or [111] direction. A more detailed analysis taking in to account anisotropies only gives quantitative corrections.

\subsubsection{Curved quantum well}

\begin{figure*}[htb]	\includegraphics{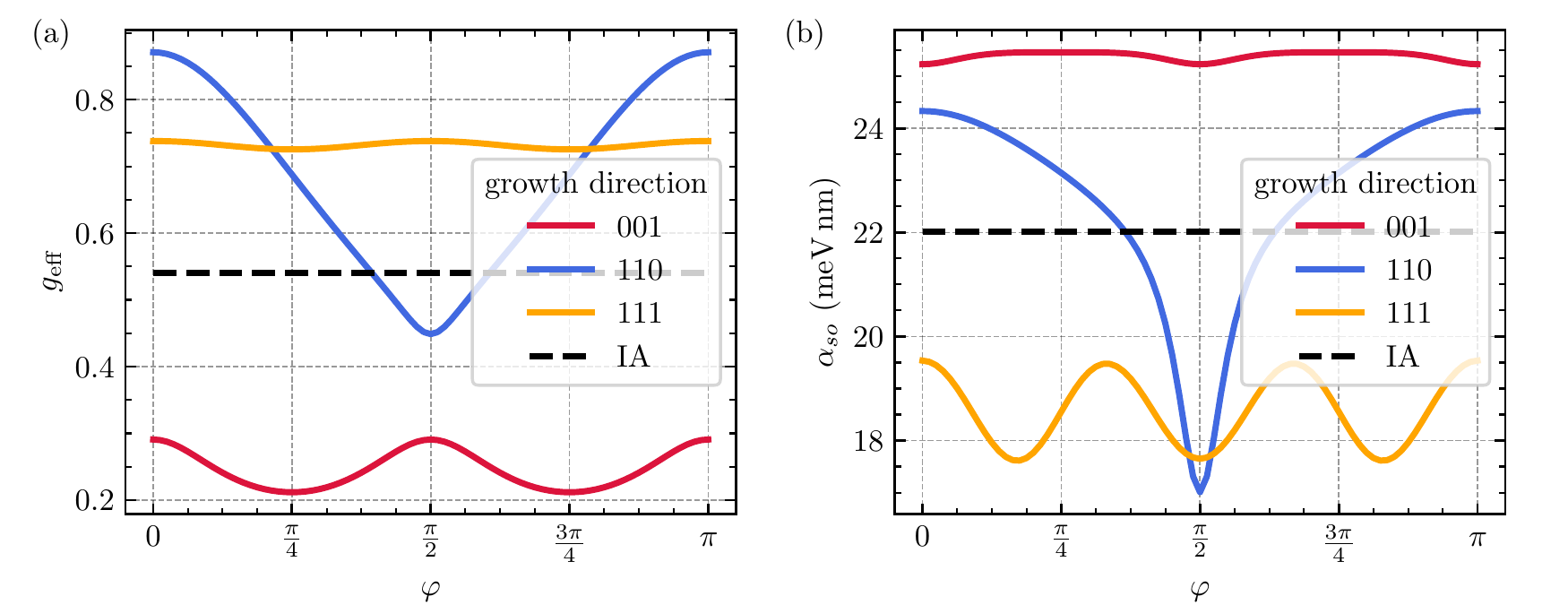}
	\caption{Numerically calculated  (a)  effective $g$ factor $g_\mathrm{eff}$ and (b) SOI $\alpha_{so}$  of a CQW with $R=\SI{15}{\nano\meter}$, $R_1 = \SI{25}{\nano\meter}$, and $R_2=\SI{35}{\nano\meter}$ [see Fig.~\ref{fig:Three1dSystems}(c)] as a function of the angle $\varphi$ that determines the direction of the electric field. We consider the three growth directions indicated by the legend and compare the results to the effective parameters calculated applying the isotropic approximation (IA) of the ILK Hamiltonian $H_\mathrm{ILK} $. The $g$ factor and the SOI strength in the isotropic case lie in between the values obtained for the different growth directions in the anisotropic case. The $g$ factor is smallest for the $[001]$ growth direction and largest for $[110]$ if, in addition, $\vect{E}\parallel x$. For the $[110]$ growth direction, the $g$ factor strongly depends on the direction of the electric field and shows a $\pi$ periodicity. In contrast, for $[111]$, the $g$ factor is almost constant. Also for the NW grown along $[001]$, the oscillations are small. The SOI is largest for the $[001]$ growth direction. Interestingly, the SOI is almost constant for the bent shell grown along $[001]$. For $[111]$, it oscillates with periodicity $\pi/3$. Similarly to the $g$ factor, the SOI for $[110]$ strongly depends on the direction of the electric field. Here, we choose $E = \SI{0.5}{\volt\per\micro\meter}$ and $B = \SI{0.1}{\tesla}$. We use a lattice spacing of \SI{0.5}{\nano\meter}. \label{fig:EffPar_B01_Rin15Rout25Rs35_str_anisotropy}}
\end{figure*}

Finally, in Fig.~\ref{fig:EffPar_B01_Rin15Rout25Rs35_str_anisotropy}, we analyze the effect of anisotropies on the properties of the CQW. As in Fig.~\ref{fig:effpar_R15_E2phi_comp}, we rotate the electric field around the $z$ direction keeping it perpendicular to the NW. Here, we fix the strength of the electric field to $E= \SI{0.5}{\volt\per\micro\meter}$ and consider the three growth directions  introduced in Sec.~\ref{sec:ModelOfNW}. We observe that the $g$ factor and the SOI strength obtained by the ILK Hamiltonian $H_\mathrm{ILK} $  are between the values estimated by using the anisotropic LK Hamiltonian. Similarly to the core/shell NW, the effective $g$ factor of the CQW is to good approximation independent of the angle $\varphi$ of the electric field for the  $z\parallel[001]$ and  $z\parallel[111]$ growth directions. The dependence on $\varphi$ is more pronounced at $z\parallel[110]$. For this growth direction and for $\vect{E}\parallel x \parallel [1\bar{1}0]$, the $g$ factor is maximal and has a value of $0.87$.
Moreover, in  $z\parallel [111]$-CQWs, the SOI oscillates between $\alpha_{so} = \SI{17.5}{\milli\electronvolt\nano\meter}$ and $\alpha_{so} = \SI{21.5}{\milli\electronvolt\nano\meter}$ with a $\pi/3$ periodicity, as expected from Ref.~\cite{Adelsberger2022}. In contrast, in the $z\parallel [001]$-CQWs  the periodicity is $\pi/2$ and  it is $\pi$ in $z\parallel [110]$-CQWs.  
We also point out that in analogy to core/shell NWs [see Fig.~\ref{fig:effpar_R15_E2phi_comp}], the amplitudes of the oscillations of $g_\mathrm{eff}$ and $\alpha_{so}$ increase at larger values of the electric field (not show here).

From these results, we conclude that the ILK Hamiltonian in CQWs provides a reasonable qualitative approximation but a more detailed analysis of the LK anisotropies is required to have a good quantitative description of the system, especially in CQWs grown along the [110] direction, where the effect of the cubic anisotropies is enhanced.

\section{Conclusions} \label{sec:conclusion}
We presented a low-energy effective model that describes holes confined in a NW with a magnetic field parallel to the NW axis and in a perpendicular electric field. 
We discussed the bulk solution for holes in isotropic semiconductors, where we include the orbital effects of the magnetic field exactly, as well as we extend this result to NWs, providing an analytical solution for holes in isotropic semiconductor NWs. These effects are found to be essential to accurately describe the properties of the NW.

In particular, we observe a strong renormalization of the effective $g$ factor due to orbital effects even at small values of the external magnetic field. By using a second-order perturbation theory, we also analyze the effects of homogeneous and inhomogeneous electric fields. The homogeneous electric field decreases the effective $g$ factor but enables a strong SOI. The average effective mass changes sign and the spin-dependent mass vanishes with increasing electric field. The inhomogeneous electric field has a strong effect at weak electric field where it leads to a decrease of the $g$ factor.
We also include strain in the system, which enhances the subband gap, thus, yielding a reduced HH-LH coupling. This effect increases the $g$ factor but decreases the SOI. 

We study also the low-energy physics of holes confined in a gate-defined one-dimensional channel and we predict a similar behavior of the $g$ factor and SOI as in a Ge NW but  with some qualitative differences in the average and spin-dependent effective mass, which are in this case only weakly dependent on the electric field. 
We also examine  holes in a CQW, where we predict a $g$ factor independent of the electric field in a wide range of parameters. This feature is relevant for spin qubits in quantum dots because it reduces the susceptibility to charge noise, a major decoherence channel in current devices. Orbital effects are also found to be extremely important in CQW, yielding an enhanced $g$ factor at weak electric fields.

We predict that the spin-orbit split-off band causes a small quantitative correction of the effective parameters in Ge NWs. For particularly thin NWs and CQWs, this effect becomes more relevant, however, the qualitative behavior remains unchanged. By a comparison of our results form calculations with the IA to results where anisotropies are taken into account we find a good agreement justifying the application of the IA.

\begin{acknowledgments}
We thank Monica Benito and Christoph Kloeffel for useful discussions and comments. 
This work is supported by the Swiss National Science Foundation (SNSF) and NCCR SPIN (grant number 51NF40-180604). 
\end{acknowledgments}

\appendix
\section{Bulk Dispersion Relation \label{sec:BulkDispersionRelation}}

In this Appendix, we show how to calculate the bulk dispersion relation for holes including orbital effects. We consider the LK Hamiltonian in Eq.~\eqref{eqn:LK_Hamiltonian}. In the symmetric gauge $\vect{A} = (-y, x, 0) B/2$, using the Landau ladder operators from Eq.~\eqref{eqn:LandauLevelOp}, the Hamiltonian explicitly reads in the spin basis $(+3/2, +1/2, -1/2, -3/2)$,
\begin{widetext}
	\begin{align}
		\frac{H_\mathrm{LK}}{\hbar\omega_c} = \begin{pmatrix}
			\frac{\gamma_-^z}{2}k_z^2 + \gamma_+ \left(a^\dagger a + \frac{1}{2}\right) & -\sqrt{6} \gamma_s a k_z & -\sqrt{3} \gamma_s a^2 & 0 \\
			 -\sqrt{6} \gamma_s a^\dagger k_z & \frac{\gamma_+^z}{2}k_z^2 + \gamma_- \left(a^\dagger a + \frac{1}{2}\right)  & 0 & -\sqrt{3} \gamma_s a^2\\
			 -\sqrt{3} \gamma_s \left(a^\dagger\right)^2 & 0 & \frac{\gamma_+^z}{2}k_z^2 + \gamma_- \left(a^\dagger a + \frac{1}{2}\right)  & \sqrt{6} \gamma_s a k_z\\
			 0 &  -\sqrt{3} \gamma_s \left(a^\dagger\right)^2 & \sqrt{6} \gamma_s a^\dagger k_z & \frac{\gamma_-^z}{2}k_z^2 + \gamma_+ \left(a^\dagger a + \frac{1}{2}\right) 
		\end{pmatrix}, \label{eqn:LKHamBulkDisp}
	\end{align}
\end{widetext}
where $\gamma_\pm^z = (\gamma_1-2\gamma_s)/2$.
Adding the Zeeman Hamiltonian from Eq.~\eqref{eqn:HamZeeman} leads to the Schr{\"o}dinger equation 
\begin{align}
	O_1(N, k_z^2)  \varphi_1 &= \sqrt{2} a k_z \varphi_2 + a^2 \varphi_3, \label{eqn:SEBulk1}\\
	O_2(N, k_z^2)  \varphi_2 &= \sqrt{2} a^\dagger k_z \varphi_1 + a^2 \varphi_4, \label{eqn:SEBulk2}\\
	O_3(N, k_z^2)  \varphi_3 &= -\sqrt{2} a k_z \varphi_4 + \left(a^\dagger\right)^2 \varphi_1, \label{eqn:SEBulk3}\\
	O_4(N, k_z^2)  \varphi_4 &= -\sqrt{2} a^\dagger k_z \varphi_3 + \left(a^\dagger\right)^2 \varphi_2, \label{eqn:SEBulk4}
\end{align}
with energies normalized by $\hbar \omega_c$, lengths normalized by $l_B$, and the components of the wave function $\varphi_i$, $i= 1,2,3,4$. The operators $O_i = O_i(N, k_z^2)$ are the diagonal entries of the Hamiltonian in Eq.~\eqref{eqn:LKHamBulkDisp} plus the Zeeman term minus the energy eigenvalue $\varepsilon$, explicitly given by 
\begin{align}
	O_1 &= \frac{1}{\sqrt{3} \gamma_s} \left(\frac{\gamma_-^z}{2} k_z^2 + \gamma_+ \left(N + \frac{1}{2}\right) - \varepsilon +\frac{3\kappa}{2}\right),\\
	O_2&= \frac{1}{\sqrt{3} \gamma_s} \left(\frac{\gamma_+^z}{2} k_z^2 + \gamma_- \left(N + \frac{1}{2}\right) - \varepsilon +\frac{\kappa}{2}\right),\\
	O_3 &= \frac{1}{\sqrt{3} \gamma_s} \left(\frac{\gamma_+^z}{2} k_z^2 + \gamma_- \left(N + \frac{1}{2}\right) - \varepsilon -\frac{\kappa}{2}\right),\\
	O_4 &= \frac{1}{\sqrt{3} \gamma_s} \left(\frac{\gamma_-^z}{2} k_z^2 + \gamma_+ \left(N + \frac{1}{2}\right) - \varepsilon -\frac{3\kappa}{2}\right),
\end{align} 
where we define the number operator $N = a^\dagger a$. 
In the following we will make use of the relation $a^\dagger a = a a^\dagger -1$, which implies $a^m O_i(N, k_z^2) = O_i(N+m, k_z^2) a^m$ and $\left(a^\dagger\right)^m O_i(N, k_z^2) = O_i(N-m, k_z^2) \left(a^\dagger\right)^m$, $m \in \mathbb{N}$. Multiplying Eq.~\eqref{eqn:SEBulk2} by $\left(a^\dagger\right)^2$ and Eq.~\eqref{eqn:SEBulk3} by $a^2$ from the left side yields
\begin{align}
	&\left(2 k_z^2 a^\dagger a + \left(a^\dagger\right)^2 a^2\right) \varphi_4\nonumber\\
	 &= -\sqrt{2} a^\dagger k_z O_3(N, k_z^2) \varphi_3+ \left(a^\dagger\right)^2 O_2(N, k_z^2) \varphi_2, \label{eqn:SEBulk2a}\\
	 &\left(2 k_z^2 a a^\dagger + a^2 \left(a^\dagger\right)^2\right) \varphi_1 \nonumber\\
	 &=  \sqrt{2} a k_z O_2(N, k_z^2) \varphi_2 + a^2 O_3(N, k_z^2) \varphi_3, \label{eqn:SEBulk3a}
\end{align}
respectively. Since the left side of these equations contains only the operators $k_z^2$ and combinations of $a$ and $a^\dagger$, it can be rewritten in terms of $N$, and thus, this part of the equations commutes with $O_i(N, k_z^2)$. Thus, applying $O_4(N, k_z^2)$ to Eq.~\eqref{eqn:SEBulk2a} and $O_1(N, k_z^2)$ to Eq.~\eqref{eqn:SEBulk3a} results in 
\begin{align}
	a^\dagger A(N+2, k_z^2) \varphi_2 &= \sqrt{2} k_z B(N+1, k_z^2) \varphi_3, \label{eqn:SEBulk2b}\\
	\sqrt{2} k_z C(N-1, k_z^2) \varphi_2 &=  - a D(N-2, k_z^2)\varphi_3, \label{eqn:SEBulk3b}
\end{align}
with
\begin{align}
	A(N, k_z^2) = &N (N + 2 k_z^2 -1) \nonumber\\
	&- O_4(N, k_z^2) O_2(N-2, k_z^2),\\
	B(N, k_z^2) = &N (N + 2 k_z^2 -1)\nonumber\\
	&- O_4(N, k_z^2) O_3(N-1, k_z^2),\\
	C(N, k_z^2) = &N^2 + 3N + 2 + 2k_z^2 (N+1)\nonumber\\
	&- O_1(N, k_z^2) O_2(N+1, k_z^2),\\
	D(N, k_z^2) = &N^2 + 3N + 2 + 2k_z^2 (N+1)\nonumber\\ 
	&- O_1(N, k_z^2) O_3(N+2, k_z^2).
\end{align}
Acting with $\sqrt{2}k_z C(N-2, k_z^2)$ on Eq.~\eqref{eqn:SEBulk2b} finally results in the implicit dispersion relation
\begin{align}
	2 k_z^2 C(\bar{n}-2, k_z^2) B(\bar{n}+1) + \bar{n} A(\bar{n}+1, k_z^2)D(\bar{n}-2, k_z^2) = 0
\end{align}
with $\bar{n}$ being the integer eigenvalue of $N$, i.e., $N \varphi_3 = \bar{n} \varphi_3$. We can solve this equation for the energies $\varepsilon$, which yields the bulk dispersion relation depicted in Fig.~\ref{fig:bulkSpec_B1}. 

We also discuss the bulk solution excluding orbital effects. In this case, we only need to diagonalize the ILK Hamiltonian given in Eq.~\eqref{eqn:LK_Hamiltonian_sphericalApp} where $H_\mathrm{orb} = 0$ and $\vect{\pi} = \vect{k}$, which can easily be done by applying the unitary transformation
\begin{align}
	U = e^{i \theta J_y} e^{i \varphi J_z}.
\end{align}
We then obtain the diagonal matrix
\begin{align}
	U H_\mathrm{LK} U^\dagger = \frac{\hbar^2}{2 m} \left(\gamma_k -2 \gamma_s J_z^2\right) k^2,
\end{align}
which yields two degenerate states with parabolic dispersion relation ($k = \abs{\vect{k}}$). The rotation angles depend only on the direction of $\vect{k}$ and are given by
\begin{align}
	\theta &= \arccos\left(\frac{k_z}{k}\right) \in [0, \pi),\\
	\varphi &= \mathrm{arctan2} \left(k_y, k_x\right) \in (-\pi, \pi],
\end{align}
with
\begin{align}
	&\mathrm{arctan2} (y, x) = \nonumber\\
    &\begin{cases}
		2 \arctan\left(\frac{y}{x+\sqrt{x^2 + y^2}}\right)\ &\mathrm{if}\ \ x>0 \|\ y \neq 0,\\
		\pi\ &\mathrm{if}\ x<0\ \wedge\ y = 0,\\
		\mathrm{undefined}\ &\mathrm{if}\ x=0\ \wedge\ y = 0.
	\end{cases}
\end{align}
At $k_x=k_y =0$ the angle $\varphi$ is undefined but, in this case, the Hamiltonian is diagonalized directly by  the rotation $e^{i \varphi J_z}$. The bulk dispersion relation without orbital effects is shown by the black solid lines in Fig.~\ref{fig:bulkSpec_B1}.

\section{Derivation of the Analytical Solution for a Cylindrical Nanowire \label{sec:AnalyticsDerivationNW}}
This Appendix provides the derivations for the exact analytical solution of an isotropic semiconductor hole NW with circular cross section in a magnetic field parallel to the NW. As a starting point, we consider the Hamiltonian for the perpendicular directions $H_{xy}$ defined in Eq.~\eqref{eqn:HamPerp} with creation and annihilation operators in polar coordinates defined as
\begin{align}
	&a^\dagger a=\frac{1}{2}\left(-\partial_r^2-\frac{1}{r}\partial_r-\frac{1}{r^2}\partial_\varphi^2+\frac{r^2}{4}-i\partial_\varphi -1 \right), \label{eqn:adga_polar}\\
	&a^\dagger=\frac{-ie^{i\varphi}}{\sqrt{2}}\left(\partial_r+\frac{i}{r}\partial_\varphi -\frac{r}{2} \right), \\
	&a=\frac{-ie^{-i\varphi}}{\sqrt{2}}\left(\partial_r-\frac{i}{r}\partial_\varphi +\frac{r}{2} \right) .
\end{align}
Also we consider now the  Zeeman Hamiltonian $H_Z$ from Eq.~\eqref{eqn:HamZeeman}. The eigenstates of the upper block ($\uparrow$) of $H_{xy}$ are eigenfunctions of $a^\dagger a$. We find the general eigenfunction of $a^\dagger a$ in Eq.~\eqref{eqn:adga_polar} by identifying the equation
\begin{align}
	a^\dagger a \left[e^{-i m \varphi} r^m e^{-r^2/4} g(\vect{r})\right] = 0
\end{align}
as the Laguerre differential equation with $m \in \mathbb{Z}$. The eigenvalues of $a^\dagger a$ are $\alpha \in \mathbb{R}$. We remark that in Sec.~\ref{sec:BulkSolution} for the bulk solution, the eigenvalues are $\alpha\to n \in \mathbb{N}$ because the bulk solutions are required to decay to zero at infinity. Here, in contrast, we have different boundary conditions, allowing for real valued $\alpha$. Most generally, the eigenfunction to the eigenvalue $\alpha$ is 
\begin{align}
	&\psi_{m, \alpha} (\vect{r}) = 2^{-\frac{m}{2}} e^{-im\varphi} e^{-\frac{r^2}{4}}r^m \nonumber\\
	&\times\left[{i^m} L^m_{\alpha}\left(\frac{r^2}{2}\right) + \frac{(-i)^m}{\alpha!} U\left(-\alpha, m+1, \frac{r^2}{2}\right)\right],
\end{align}
where $L_{a}^b \left(x\right)$ is the associated Laguerre function and $U(a, b, x)$ is the confluent hyper-geometric function of the second kind. The effect of the creation and annihilation operators on this eigenfunction is given by
\begin{align}
&	a^\dagger \psi_{m,\alpha}(\textbf{r})=(1+\alpha)\psi_{m-1,\alpha+1}(\textbf{r}),\\
&	a \psi_{m,\alpha}(\textbf{r})=\psi_{m+1,\alpha-1}(\textbf{r}).
\end{align}
The general eigenstate of the $\uparrow$ block of $H_{xy}$ in Eq.~\eqref{eqn:HamPerp} is of the form $\left(\psi_{m,\alpha}(\textbf{r}),c^\uparrow \psi_{m-2,\alpha+2}(\textbf{r})\right)$, which we can insert into the Schr{\"o}dinger equation given by the $\uparrow$ block of $H_{xy}$ in Eq.~\eqref{eqn:HamPerp}, arriving at
	\begin{align}
		&\frac{1}{\sqrt{3}\gamma_s}\left[(\gamma_1+\gamma_s) \left(\alpha + \frac{1}{2} \right)-\varepsilon_{+3/2}\right] \psi_{m, \alpha} (\vect{r})\nonumber\\
		& = c^\uparrow\psi_{m, \alpha} (\vect{r}), \label{eqn:schrodinger1}\\
		&\frac{c^\uparrow}{\sqrt{3}\gamma_s}\left[(\gamma_1-\gamma_s) \left(\alpha + \frac{5}{2} \right)-\varepsilon_{-1/2}\right] \psi_{m-2, \alpha+2} (\vect{r}) \nonumber\\
		&= (\alpha +1)(\alpha+2) \psi_{m-2, \alpha+2} (\vect{r}), \label{eqn:schrodinger2}
	\end{align}
with the energies $\varepsilon_{+3/2} = \varepsilon - 3\kappa/2 $ and $\varepsilon_{-1/2} = \varepsilon +\kappa/2$ redefined to include the Zeeman energy. The coefficients are given in the main text by Eqs.~\eqref{eqn:alphaU} and~\eqref{eqn:cU}.

By imposing HW boundary conditions, following from Eq.~\eqref{eqn:HW_boundaryCondition}, and by requiring each element of the spinor to vanish at $r=R$, we  arrive at the expression for the wave function given in Eq.~\eqref{eqn:WF_GeNWU}. For the lower block ($\downarrow$) of $H_{xy}$ in Eq.~\eqref{eqn:HamPerp} describing the spin states ($-3/2$, $+1/2$) we proceed analogously with a similar ansatz $\left(\psi_{m,\alpha}(\textbf{r}),  c^\downarrow \psi_{m-2,\alpha+2}(\textbf{r})\right)$ as for the upper block ($\uparrow$). We obtain in this case the coefficients given by Eqs.~\eqref{eqn:alphaD} and~\eqref{eqn:cD}. With the ansatz for the $\downarrow$ block,  we arrive at the spinor given in Eq.~\eqref{eqn:WF_GeNWD}.

It is possible to include strain into our analytical calculations because of the simple form of the BP Hamiltonian $H_{\rm BP}$ defined in Eq.~\eqref{eqn:BPHam}. Since $H_{\rm BP} \propto J_{z'}^2$ does not change the Schr{\"o}dinger in Eqs.~\eqref{eqn:schrodinger1} and~\eqref{eqn:schrodinger2} qualitatively the calculation of the eigenstates is analogous. The solution keeps the same form, however, the coefficients $c_{\pm}^s$ and $\alpha_{\pm}^s$ are modified as
\begin{widetext}
	\begin{align}
		c_{\pm}^\uparrow= &\frac{(2 \alpha_{\pm}^\uparrow +1) (\gamma_1+ \gamma_s)+3\kappa -2 \varepsilon }{2 \sqrt{3} \gamma_s} + \frac{9\abs{b}\varepsilon_s}{4\sqrt{3}\gamma_s \hbar \omega_c},  \label{eqn:cUs}\\
		c_{\pm}^\downarrow= &\frac{(2 \alpha_{\pm}^\downarrow +1) (\gamma_1- \gamma_s)+\kappa -2 \varepsilon }{2 \sqrt{3} \gamma_s} + \frac{\abs{b}\varepsilon_s}{4\sqrt{3}\gamma_s \hbar \omega_c},  \label{eqn:cDs}\\
		\alpha_\pm^\uparrow = &\frac{- 6\gamma_1^2+24\gamma_s^2+4 \gamma_1\varepsilon-2\gamma_1\kappa+4\gamma_s\kappa+(-5\gamma_1+4\gamma_s)\frac{\abs{b}\varepsilon_s}{\hbar\omega_c}}{4(\gamma_1^2-4\gamma_s^2)}\nonumber\\
		&+ \Bigg\{4\gamma_1^4 - 23\gamma_1^2 \gamma_s^2 + 28\gamma_s^4 - 8\gamma_1^2\gamma_s\varepsilon + 32\gamma_s^3\varepsilon + 16\gamma_s^2 \varepsilon^2 - 4(\gamma_1-2\gamma_s)\left[2\gamma_1^2 + 3\gamma_1\gamma_s - 2\gamma_s(\gamma_s+\varepsilon)\right]\kappa \nonumber \\
		&+ 4(\gamma_1-2\gamma_s)(\gamma_1+\gamma_s)\kappa^2 - 2 \left\{4\gamma_1^3 + 2\gamma_s^2\left[10(\gamma_s+\varepsilon)+\kappa\right] -  \gamma_1^2(5\gamma_s+4\kappa) + \gamma_1 \gamma_s \left[-4(4\gamma_s+\varepsilon)+7\kappa\right] \right\}\frac{\abs{b}\varepsilon_s}{\hbar\omega_c}\nonumber\\
		&+\left(4\gamma_1^2+10\gamma_1\gamma_s+13\gamma_s^2\right)\left(\frac{\abs{b}\varepsilon_s}{\hbar\omega_c}\right)^2\Bigg\}^{1/2}/\left[2(\gamma_1^2-4\gamma_s^2)\right],\label{eqn:alphaUs}\\
		\alpha_\pm^\downarrow = &\frac{- 6\gamma_1^2+24\gamma_s^2+4 \gamma_1\varepsilon+2\gamma_1\kappa-4\gamma_s\kappa+(-5\gamma_1+4\gamma_s)\frac{\abs{b}\varepsilon_s}{\hbar\omega_c}}{4(\gamma_1^2-4\gamma_s^2)}\nonumber\\
		&+ \Bigg\{4\gamma_1^4 - 23\gamma_1^2 \gamma_s^2 + 28\gamma_s^4 + 8\gamma_1^2\gamma_s\varepsilon + 32\gamma_s^3\varepsilon + 16\gamma_s^2 \varepsilon^2 - 4(\gamma_1-2\gamma_s)\left[2\gamma_1^2 + 3\gamma_1\gamma_s + 2\gamma_s(-\gamma_s+\varepsilon)\right]\kappa \nonumber \\
		&+ 4(\gamma_1-2\gamma_s)(\gamma_1+\gamma_s)\kappa^2 + 2 \left\{4\gamma_1^3 + 2\gamma_s^2\left[10(\gamma_s-\varepsilon)+\kappa\right] -  \gamma_1^2(5\gamma_s+4\kappa) + \gamma_1 \gamma_s \left[4(-4\gamma_s+\varepsilon)+7\kappa\right] \right\}\frac{\abs{b}\varepsilon_s}{\hbar\omega_c}\nonumber\\
		&+\left(4\gamma_1^2-10\gamma_1\gamma_s+13\gamma_s^2\right)\left(\frac{\abs{b}\varepsilon_s}{\hbar\omega_c}\right)^2\Bigg\}^{1/2}/\left[2(\gamma_1^2-4\gamma_s^2)\right].\label{eqn:alphaDs}
	\end{align}
\end{widetext}
For $b=0$ the coefficients coincide with the ones given in the main text by Eqs.~\eqref{eqn:alphaU}, ~\eqref{eqn:cU}, ~\eqref{eqn:alphaD}, and~\eqref{eqn:cD}.

\section{Effective Parameters \label{sec:effParam}}

\begin{figure*}[htb]
	\includegraphics{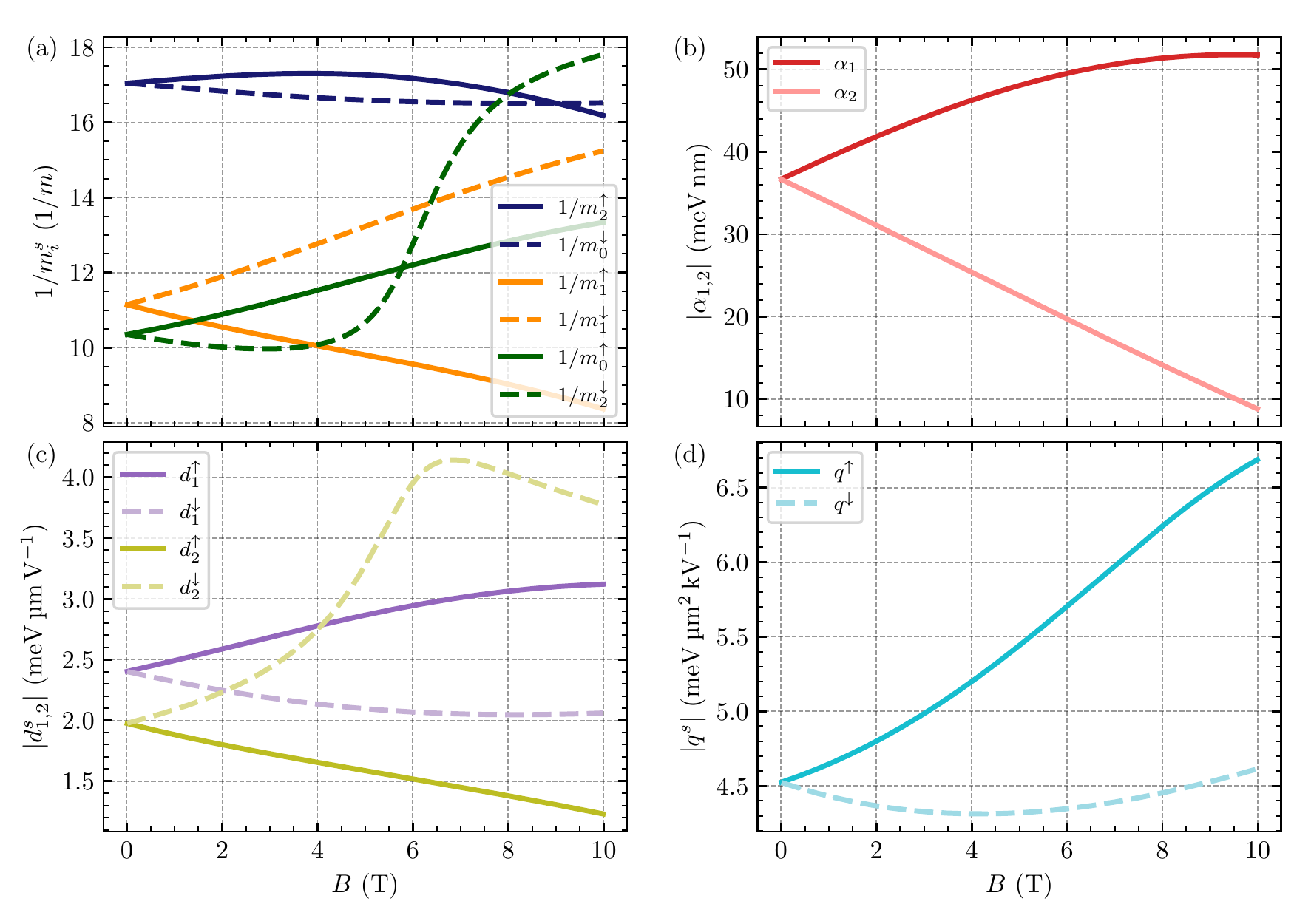}
	\caption{Effective parameters of a Ge NW with circular cross section of radius $R=\SI{15}{\nano\meter}$ as a function of $B$ calculated semi-analytically at $E_x = E_y = \delta E = 0$. The solid lines in (a), (c), and (d) correspond to states from the $\uparrow$ subspace and the dashed ones to the $\downarrow$ subspace. (a) The effective masses $1/m_i^s$ are calculated by using Eq.~\eqref{eqn:GEmeff}. The masses of the lowest two energy states, $m_2^\uparrow$ and $m_0^\downarrow$ (blue) depend only weakly on $B$ compared to the masses of the next higher in energy states, $m_1^\uparrow$ and  $m_1^\downarrow$ (orange) or $m_0^\uparrow$ and $m_2^\downarrow$ (green). The mass $m_2^\downarrow$ (green dashed) varies strongly with $B$ due to the anti-crossing in the spectrum [cf. Fig.~\ref{fig:gFacRen}(a)]. The colors of the lines are the same as for the corresponding states in Fig.~\ref{fig:gFacRen}(a). (b) The SOI term $|\alpha_1|$ [cf. Eq.~\eqref{eqn:SOcouplingGe1}] increases with $B$ and reaches a maximum at $B=\SI{9.5}{\tesla}$, while $|\alpha_2|$ [cf. Eq.~\eqref{eqn:SOcouplingGe2}] decreases linearly. (c) The dipole moments $|d_{1,2}^{\rm{U,D}}|$ are given by Eqs.~\eqref{eqn:dipCoup1} and~\eqref{eqn:dipCoup2}. For our choice of parameters, $|d_1^\uparrow|$ (violet solid) increases monotonically with $B$, $|d_1^\downarrow|$ (violet dashed) has a minimum, $|d_2^\uparrow|$ (green solid) decreases monotonically, and $|d_2^\downarrow|$  (green dashed) exhibits a distinctive maximum [cf. the anti-crossing in Fig.~\ref{fig:gFacRen}(a)]. (d) The quadruple moments  $q^s$ are given by Eq.~\eqref{eqn:quadCoup1}. The quadrupole moment $|q^\uparrow|$ increases monotonically and  $|q^\downarrow|$ increases after reaching a minimum at $B=\SI{4.7}{\tesla}$.  Note that in panels (b), (c), and (d) we present absolute values. \label{fig:effParGe_R15_str0}}
\end{figure*}

The Hamiltonian $H_\mathrm{LK}^\mathrm{eff}$ introduced in Eq.~\eqref{eqn:effHam} in Sec.~\ref{sec:EffHam} depends on several effective parameters that are defined in the main text. In this Appendix we show the dependence of these parameters (cf. Fig.~\ref{fig:effParGe_R15_str0}) on the magnetic field. These parameters enter the calculations for the $2\times 2$ NW Hamiltonian in Sec.~\ref{sec:2x2wireHam}.

The inverse effective masses $1/m_i^s$ are shown in Fig.~\ref{fig:effParGe_R15_str0}(a).  The ground state at $B=0$ is almost exclusively of LH nature. Therefore, we would expect the inverse ground state masses to be $1/m_2^\uparrow = 1/m_0^\downarrow = \gamma_1+2\gamma_s = 23.3/m$ at $B=0$, see Eq.~\eqref{eqn:GEmeff}. However, the corrections from second-order perturbation theory are large, yielding a larger mass. The states higher in energy are a mixture of HH and LH, and therefore, their masses are larger. Also, these values are considerably corrected by second-order perturbation theory. The mass $m_2^\downarrow$ decreases strongly above $B=\SI{5}{\tesla}$, a trend that can be explained by the anticrossing of states in the spectrum in Fig.~\ref{fig:gFacRen}(a). There is a clear avoided crossing between the green-dashed and the brown-dashed lines.

In Fig.~\ref{fig:effParGe_R15_str0}(b), we plot the absolute value of the two SOI parameters $\alpha_1$ and $\alpha_2$ obtained from Eq.~\eqref{eqn:SOcouplingGe1} and Eq.~\eqref{eqn:SOcouplingGe2} as a function of the magnetic field $B$. Both parameters exhibit a linear dependence on the magnetic field for weak fields. Moreover, $\alpha_2$ has a linear behavior in the whole range of $B$ shown, while $\alpha_1$ reaches a maximum at $B=\SI{9.5}{\tesla}$. At $B=0$ both couplings are $\alpha_{1/2}(B=0) \approx 2.5\, \hbar^2/(m R)$; at larger values of $B$, $\alpha_1$ increases with the magnetic field while $\alpha_2$ decreases.

The behaviors of the absolute values of the dipole and the quadrupole moments in a NW of radius $R=\SI{15}{\nano\meter}$ are analyzed in Figs.~\ref{fig:effParGe_R15_str0}(c) and~\ref{fig:effParGe_R15_str0}(d), respectively. The dipole moment $|d_1^\downarrow|$ has a  minimum at $B=\SI{8.5}{\tesla}$ and  $|d_1^\uparrow|$ increases throughout the whole magnetic field range; moreover  $|d_2^\uparrow|$ decreases linearly and $|d_2^\downarrow|$ has a  maximum at $B=\SI{6.7}{\tesla}$  where it reaches $\SI{4.2}{\milli\electronvolt\micro\meter\per \volt}$. The strong magnetic field dependence of $|d_2^\downarrow|$ results from the already discussed anticrossing in the spectrum in Fig.~\ref{fig:gFacRen}(a).

The quadrupole moment $|q^\uparrow|$ strongly increases with the magnetic field while $|q^\downarrow|$ only weakly depends on $B$, exhibiting a minimum at $B=\SI{4.7}{\tesla}$. In general, there are terms originating from the quadrupole moments also in the diagonal part of $	H_\mathrm{LK}^\mathrm{eff}$ given by Eq.~\eqref{eqn:effHam}. However, they are negligible compared to the contributions form $H_{xy}$, $H_Z$, and the mass terms, and we do not consider them here.  As can be seen from the Hamiltonian  $	H_\mathrm{LK}^\mathrm{eff}$ [see Eq.~\eqref{eqn:effHam}], the homogeneous field couples states from equal subspaces with neighboring $m$, while the field gradient couples the zero magnetic field ground states to the second excited states of equal subspaces. We note that the matrix elements for the dipole moments $d_{1,2}^s$ are imaginary numbers whereas the quadrupole moments $q^s$ are real numbers.

\section{Effective Model\label{sec:effMod}}
In this Appendix, we give more details on the calculations for the effective $2\times 2$ model Hamiltonian $H^{2\times 2} $ from Sec.~\ref{sec:EffHam}. The definitions introduced in Eqs.~\eqref{eqn:enDiffU}--\eqref{eqn:orbEn} allow us to write the $4\times 4$-matrix $H^{4\times 4}$ spanned by the four component basis $(\varphi^\uparrow_2, \varphi^\uparrow_1, \varphi^\downarrow_0, \varphi^\downarrow_1)$ as
\begin{widetext}
	\begin{align}
		H^{4\times 4} = 
		\begin{pmatrix}
			\varepsilon^\uparrow_+ + \Omega^\uparrow \cos(2 \theta^\uparrow)+\frac{\hbar^2 k_z^2}{2 m_2^\uparrow} & i \Omega^\uparrow \sin(2 \theta^\uparrow)  & 0 & \alpha_1 k_z\\
			-i \Omega^\uparrow \sin(2 \theta^\uparrow)  & \varepsilon^\uparrow_+ - \Omega^\uparrow \cos(2 \theta^\uparrow) +\frac{\hbar^2 k_z^2}{2 m_1^\uparrow} & \alpha_2 k_z  & 0\\
			0 & \alpha_2 k_z & \varepsilon^\downarrow_+ + \Omega^\downarrow \cos(2 \theta^\downarrow) +\frac{\hbar^2 k_z^2}{2 m_0^\downarrow}& i \Omega^\downarrow \sin(2 \theta^\downarrow) \\
			\alpha_1 k_z & 0 & - i \Omega^\downarrow \sin(2 \theta^\downarrow) &  \varepsilon^\downarrow_+ - \Omega^\downarrow \cos(2 \theta^\downarrow) +\frac{\hbar^2 k_z^2}{2 m_1^\downarrow}
		\end{pmatrix}.
	\end{align}
\end{widetext}
In order to treat the electric field exactly, we diagonalize the block in the upper left and the lower right of this matrix at $k_z=0$ via the unitary transformation $U^{-1} H^{4\times 4} U$ where 
\begin{align}
	U =
	\begin{pmatrix}
		\cos(\theta^\uparrow) & - i \sin(\theta^\uparrow) & 0 & 0 \\
		- i \sin(\theta^\uparrow)   & \cos(\theta^\uparrow) & 0 & 0\\
		0 & 0 & \cos(\theta^\downarrow) & - i \sin(\theta^\downarrow)\\
		0 & 0 & -i \sin(\theta^\downarrow) &\cos(\theta^\downarrow)
	\end{pmatrix}.
\end{align}
We use the resulting matrix and perform a second-order perturbation theory with respect to the ground state subspace $(\varphi^\uparrow_2, \varphi^\downarrow_0)$ and  we project the results onto this subspace, yielding the effective $2\times 2$ model in Eq.~\eqref{eqn:Heff2x2}.

\section{Six-Band Luttinger-Kohn Model \label{sec:6bandLKHam}}

 In addition to the four HH and LH states considered in the Hamiltonian $H$ defined  in Eq.~\eqref{eqn:FullHamiltonian}, we include in our calculations also the SOB. In this case, we use the following Hamiltonian~\cite{Winkler2003, Ahn1995, Luttinger1955, Luttinger1956}:
\small
\begin{align}
	&H^{6\times 6} = \nonumber\\
	&\begin{pmatrix}
		P+Q & S & R & 0 & -\frac{1}{\sqrt{2}} S& - \sqrt{2} R \\
		S^\ast & P-Q & 0 & R  & \sqrt{2} Q & \sqrt{\frac{3}{2}} S\\
		R^\ast & 0 & P-Q & -S & \sqrt{\frac{3}{2}} S^\ast & -\sqrt{2} Q\\
		0 & R^\ast & - S^\ast & P+Q & \sqrt{2} R^\ast & -\frac{1}{\sqrt{2}} S^\ast \\
		-\frac{1}{\sqrt{2}} S^\ast & \sqrt{2} Q & \sqrt{\frac{3}{2}} S & \sqrt{2} R & P + \Delta_{SO} & 0 \\
		- \sqrt{2} R^\ast & \sqrt{\frac{3}{2}} S^\ast & -\sqrt{2} Q & -\frac{1}{\sqrt{2}} S & 0 & P + \Delta_{SO}
	\end{pmatrix}, \label{eqn:LKHam6x6}
\end{align}\normalsize
where the matrix entries are defined as 
\begin{align}
	P &= \frac{\hbar^2}{2 m} \gamma_1 \vect{k}^2,\\
	Q &= \frac{\hbar^2}{2 m} \gamma_2 (k_x^2 + k_y^2 - 2k_z^2),\\
	S &= -\frac{\hbar^2}{2 m} 2\sqrt{3} \gamma_3 k_z k_-,\\
	R &= -\frac{\hbar^2}{2 m} \frac{\sqrt{3}}{2}\left[(\gamma_2+\gamma_3) k_-^2+(\gamma_2-\gamma_3)k_+^2\right],
\end{align}
with $k_\pm = k_\mp^\ast = k_x \pm i k_y$. The spin-orbit gap for Ge is $\Delta_{SO} =\SI{296}{\milli\electronvolt}$. Including the SOB, the Zeeman Hamiltonian $H_Z$ from Eq.~\eqref{eqn:HamZeeman} becomes~\cite{Winkler2003}
\small
\begin{align}
	&H_Z^{6 \times 6} =  \kappa \mu_B \times \nonumber\\
	&\begin{pmatrix}
		3 B_z & \sqrt{3} B_- & 0  & 0 & -\sqrt{\frac{3}{2}} B_-& 0 \\
		\sqrt{3} B_+ & B_z & B_- & 0 & \sqrt{2} B_z & -\frac{1}{\sqrt{2}} B_- \\
		0 & B_+ & - B_z & 	\sqrt{3} B_- & \frac{1}{\sqrt{2}} B_+& \sqrt{2} B_z \\
		0 & 0 & 	\sqrt{3} B_+ & -3 B_z & 0 &  \sqrt{\frac{3}{2}} B_+ \\
		-\sqrt{\frac{3}{2}} B_+ & \sqrt{2} B_z & \frac{1}{\sqrt{2}} B_- & 0 & B_z & B_- \\
		0 & - \frac{1}{\sqrt{2}} B_+ & \sqrt{2} B_z & \sqrt{\frac{3}{2}} B_- & B_+ & -B_z
	\end{pmatrix}, \label{eqn:ZeemanHam6x6}
\end{align}
\normalsize
where we define $B_\pm = B_x \pm i B_y$. 

The BP Hamiltonian including the SOB in the Ge/Si core/shell NW is given by
\begin{align}
	H_\mathrm{BP}^{6\times 6} = \abs{b}\varepsilon_s \begin{pmatrix}
		\frac{9}{4} & 0 & 0  & 0 &0 & 0 \\
	    0 & \frac{1}{4} &0 & 0 & \sqrt{2} & 0 \\
		0 & 0 & \frac{1}{4} & 0 & 0 & -\sqrt{2}\\
		0 & 0 & 0 & \frac{9}{4}& 0 &  0 \\
		0 & \sqrt{2} & 0 & 0 & \frac{5}{4} & 0\\
		0 & 0 & -\sqrt{2} & 0 & 0 & \frac{5}{4}
	\end{pmatrix},
\end{align}
while in a CQW it is given by ~\cite{Bosco2022}
\begin{widetext}
\begin{align}
	H_\mathrm{BP}^{6\times 6, \mathrm{CQW}} = \abs{b}
	\begin{pmatrix}
		\frac{9}{8} (\varepsilon_r + 2 \varepsilon_z) & 0 & -\frac{\sqrt{3}}{2} e^{-2 i \theta} \varepsilon_r & 0 &0 & \sqrt{\frac{3}{2}} e^{-2 i \theta} \varepsilon_r \\
		0 & \frac{1}{8} (\varepsilon_r + 2 \varepsilon_z) &0 & -\frac{\sqrt{3}}{2} e^{-2 i \theta} \varepsilon_r & \frac{\varepsilon_r+2\varepsilon_z}{\sqrt{2}}  & 0 \\
		-\frac{\sqrt{3}}{2} e^{2 i \theta} \varepsilon_r & 0 & \frac{1}{8} (\varepsilon_r + 2 \varepsilon_z) & 0 & 0 & -\frac{\varepsilon_r+2\varepsilon_z}{\sqrt{2}} \\
		0 & -\frac{\sqrt{3}}{2} e^{2 i \theta} \varepsilon_r & 0 &\frac{9}{8} (\varepsilon_r + 2 \varepsilon_z)& -\sqrt{\frac{3}{2}} e^{2 i \theta} \varepsilon_r &  0 \\
		0 & \frac{\varepsilon_r+2\varepsilon_z}{\sqrt{2}}  & 0 & -\sqrt{\frac{3}{2}} e^{-2 i \theta} \varepsilon_r & \frac{5}{8} (\varepsilon_r + 2 \varepsilon_z) & 0\\
		\sqrt{\frac{3}{2}} e^{2 i \theta} \varepsilon_r & 0 & -\frac{\varepsilon_r+2\varepsilon_z}{\sqrt{2}} & 0 & 0 & \frac{5}{8} (\varepsilon_r + 2 \varepsilon_z)
	\end{pmatrix} \label{eqn:6x6BPHam}
\end{align}
\end{widetext}
Here, we introduce the polar coordinate angle $\theta$ (cf. Sec.~\ref{sec:thinShell}) and the strain energies are defined in Eqs.~\eqref{eqn:strainLong} and~\eqref{eqn:strainRad}.

\clearpage

\bibliography{Literature}

\end{document}